\documentclass[aas2pp4,natbib209]{emulateapj}

\usepackage{epstopdf}

\newcommand{\Msun}{\mathrm{M}_{\sun}}

\slugcomment{To appear in ApJ}

\begin{document}


\title{Circumbinary Ring, Circumstellar disks and accretion in the binary system UY Aurigae}.

\author{Ya-Wen Tang$^1$, Anne Dutrey$^2$, St\'ephane Guilloteau$^2$, Vincent Pietu$^3$, Emmanuel Difolco$^2$, Tracy Beck$^4$, Paul T. P. Ho$^1$, Yann Boehler$^5$,
 Fr\'ederic Gueth$^3$,  Jeff Bary$^6$, Michal Simon$^7$}

\affil{Academia Sinica, Institute of Astronomy and Astrophysics, Taipei, Taiwan}
\affil{Universit\'e de Bordeaux, Observatoire Aquitain des Sciences de l'Univers,
CNRS, UMR 5804, Laboratoire d'Astrophysique de Bordeaux,
2 rue de l'Observatoire, BP 89, F-33271 Floirac Cedex, France}
\affil{IRAM, 300 rue de la piscine, F-38406 Saint Martin d'H\`eres Cedex, France}
\affil{Space Telescope Science Institute,3700 san Martin Dr. Baltimore, MD 21218, USA}
\affil{Centro de Radioastronom\`ia y Astrof\`isica, UNAM, Apartado Postal 3-72, 58089 Morelia, Michoac\`an, Mexico}
\affil{Department of Physics and Astronomy, Colgate University, 13 Oak Drive, Hamilton, NY 13346, USA}
\affil{Stony Brook University, Stony Brook, NY 11794-3800, USA}

\email{ywtang@asiaa.sinica.edu.tw}

\begin{abstract}
Recent exo-planetary surveys reveal that planets can orbit and survive around binary stars. This suggests that some fraction of young binary systems which possess massive circumbinary disks (CB) may be in the midst of planet formation.
However, there are very few CB disks detected. 
We revisit one of the known CB disks, the UY Aurigae system, and probe $^{13}$CO 2-1, C$^{18}$O 2-1, SO 5(6)-4(5) and $^{12}$CO 3-2 line emission and the thermal dust continuum. 
Our new results confirm the existence of the CB disk.
In addition, the circumstellar (CS) disks are clearly resolved in dust continuum at 1.4 mm.
The spectral indices between the wavelengths of 0.85 mm and 6 cm are found to be surprisingly low, being 1.6 for both CS disks.
The deprojected separation of the binary is 1$\farcs$26 based on our 1.4 mm continuum data. This is 0$\farcs$07 (10 AU) larger than in earlier studies.
Combining the fact of the variation of UY Aur B in $R$ band, we propose that the CS disk of an undetected companion UY Aur Bb obscures UY Aur Ba.
A very complex kinematical pattern inside the CB disk is observed due to a mixing of Keplerian rotation of the CB disk, the infall and outflow gas.
The streaming gas accreting from the CB ring toward the CS disks and possible outflows are also identified and resolved.
The SO emission is found to be at the bases of the streaming shocks.
Our results suggest that the UY Aur system is undergoing an active accretion phase from the CB disk to the CS disks.
The UY Aur B might also be a binary system, making the UY Aur a triple system.
\end{abstract}

\section{Introduction}
A substantial fraction of stars form in binary or multiple systems \citep{Duchene+2013}, and many recent results of exo-planetary surveys reveal that planets
can orbit and survive both around the individual components of a binary star and
in circumbinary (CB) orbits \citep{Udry+2007,Roell+2012}. This suggests that at least a reasonable fraction of young binary systems exhibit
circumstellar (CS) disks and CB rings. 
To be gravitationally stable, the CS disks are expected to be located inside the Roche lobe, and the CB rings are expected to be outside the outer Lindblad resonances \citep[or the 4:1 resonance for a binary with low to moderate eccentricity;][]{Artymowicz1991}.  
Any residual material inside the non stable intermediate area in between the CB rings and CS disks would inflow toward the CS disks through streamers.
These streamers feed the CS disks, which otherwise would dissipate within a relatively short time scale. 
Signs of accretion and Near InfraRed (NIR) excesses
are observed around many T Tauri binary systems of a few million years. 
Thus, streamers are expected to be common and to provide
enough material from the outer CB disk to maintain the inner CS disks, to sustain accretion onto stars and (maybe) to form planets.
Only a few sources have been observed with a clear signature of streamers. The best evidence in binary systems remains the infrared image of SR 24 \citep{Mayama+2010}, although the kinematics and mass
of the infalling material were not directly measured.

With the exception of the GG Tau ring \citep{dutrey1994} and of CB disk around V4046 Sgr \citep{Rosenfeld+2013},
the number of CB disks detected so far is also surprisingly small, partly because of the limited sensitivity of mm/submm arrays prior ALMA.
The formation and survival of a fully shaped CB disk requires very specific physical conditions.
(M)HD simulations show that this depends on the initial angular momentum, the binary mass ratio \citep{Bate1997} and the strength
and distribution of the magnetic field in the initial molecular core  \citep{Machida+2008}. In fact, spiral arms form, and they more or less mimic
a ring shape \citep[for illustration, see Figure 5 from][]{Machida+2008} or \citep[Figure 2 from][]{Bate1997}. 
This is particularly true if the brightness
contrast between the main spiral arm and the others is significant (10 or even more) under moderate angular resolutions with respect to the width of 
the main spiral arm. 
Thanks to the resolving power and high contrast of new NIR facilities, spiral patterns have been resolved around many young systems recently \citep{Fukagawa+2010}.
They are also seen in recent images obtained from mm/submm arrays such as those of the disks orbiting MW758 \citep{Isella+etal_2010} or AB Aurigae
\citep{Tang+etal_2012}. In most cases, more observations (with an adequate hydrodynamical modeling) are needed to fully characterize the behavior
and origin of such dynamical disturbances. 

To study the origin and evolution of accretion processes in young multiple T Tauri systems, 
we have chosen to focus our interest on UY Aurigae (hereafter, UY Aur). 
UY Aur is among the few binary which exhibits a ring-like CB disk \citep{Close+etal_1998}. 
It is a system separated by 0$\farcs$88 in the Taurus-Aurigae star forming region.
We assume UY Aur is at a distance of 140 pc.
Based on the spectral energy distribution, the UY Aur stellar components are classified as class II objects \citep{Beckwith+etal_1990}.
The primary, UY Aur A, has a spectral type K7 (0.95 M$_{\sun}$), and the secondary, UY Aur B, has a spectral type M0 (0.6 M$_{\sun}$) based on the photometric absorption features \citep{Herbst+etal_1995}.
A more recent estimate of the binary properties, based on the spectroscopy survey by \citet{Hartigan2003}, 
provides a spectral type of  M0 (0.6 M$_{\sun}$) and M2.5 (0.34 M$_{\sun}$) for UY Aur A and B, respectively.
%
\begin{figure*}[ht!]
\begin{center}
\includegraphics[scale=0.3]{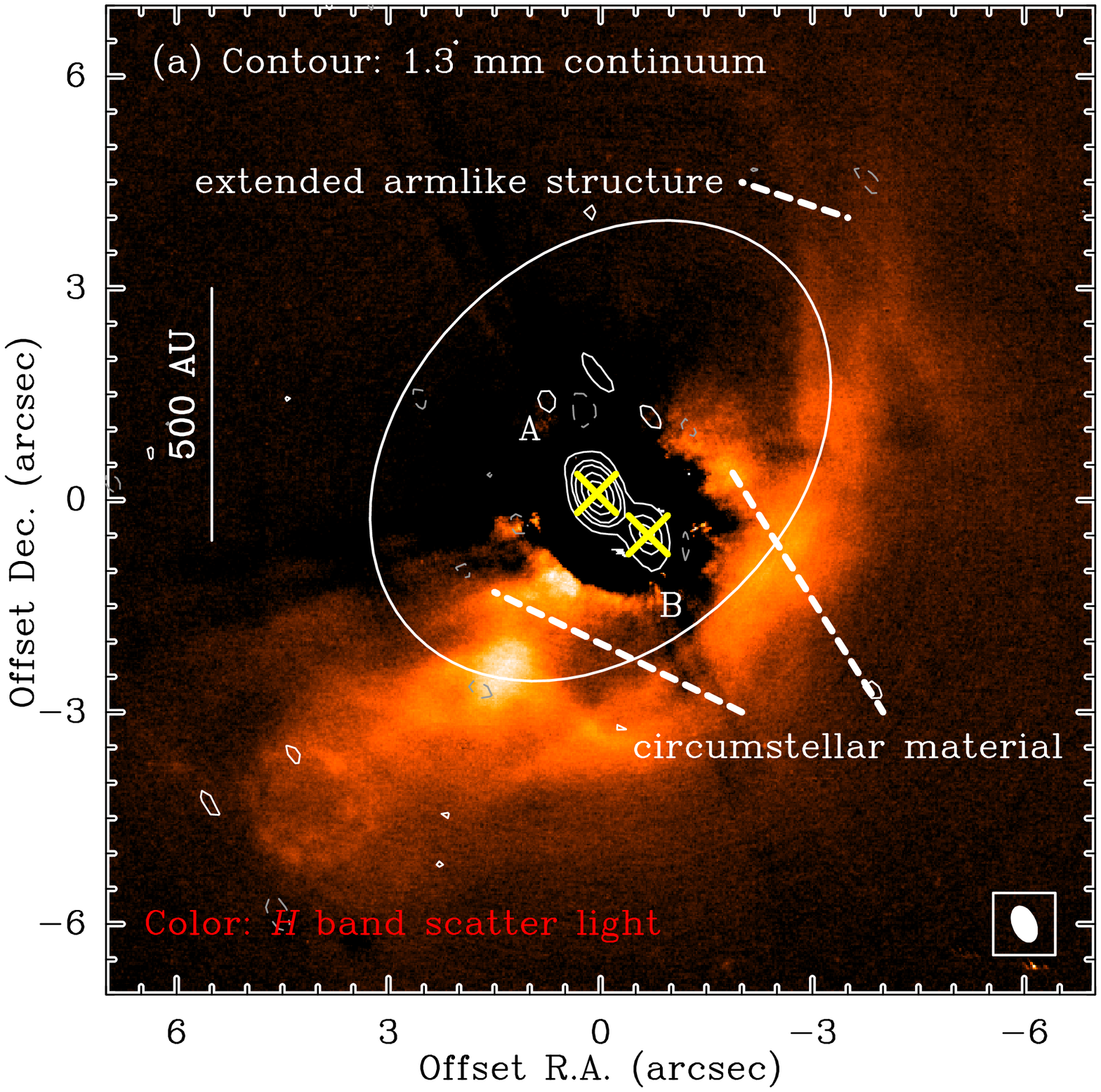}
\includegraphics[scale=0.3]{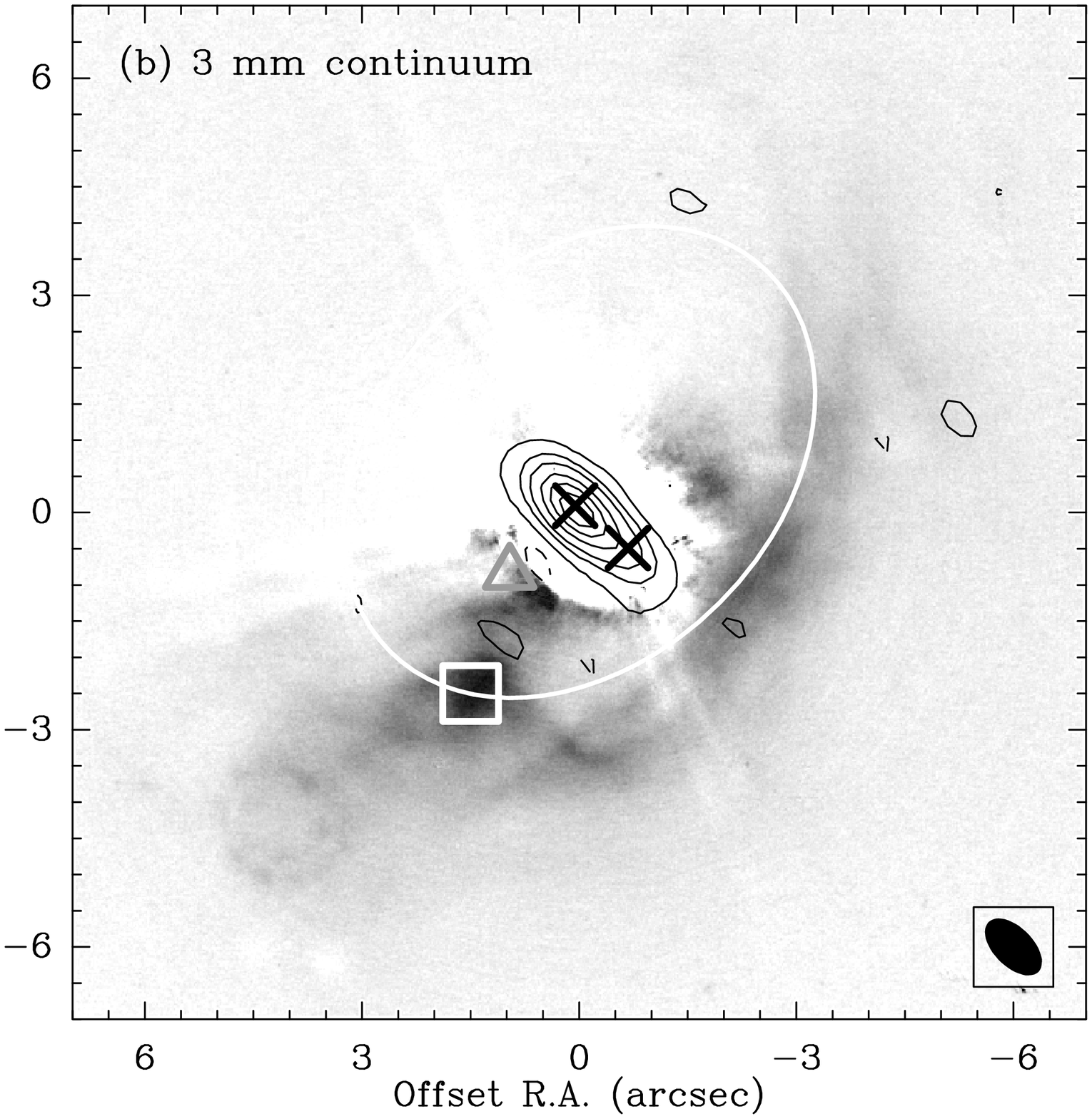}
\includegraphics[scale=0.3]{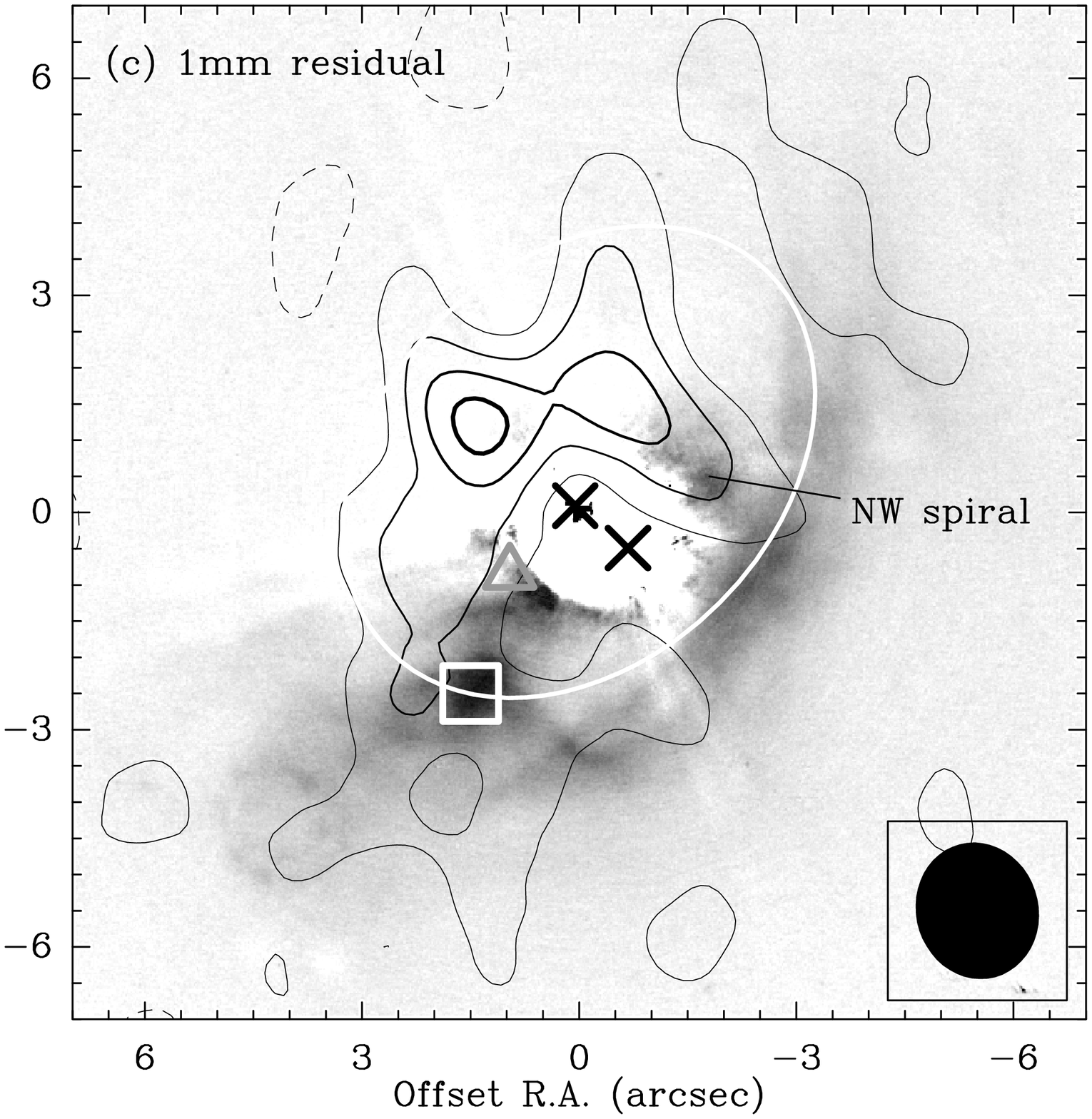}
\caption{Panel (a): Continuum emission at 1.4 mm in white contours at levels 1, 4, 7, 15, 25 $\times$ 0.54 (3$\sigma$) mJy per 0$\farcs$55 $\times$ 0$\farcs$34 with position angle (P.A.) 24$\degr$ beam.
Panel (b): Continuum emission at 3 mm in black contours at 1, 5, 10, 20, 30, 40 $\times$ 0.11 (3$\sigma$) mJy per 0$\farcs$97$\times$0$\farcs$54 beam with P.A. 45$\degr$.
Panel (c): Residual continuum emission after removing two gaussian fit at UY Aur A and B. Contour steps are 3$\sigma$ (0.323 mJy per 1$\farcs$9$\times$1$\farcs$7 beam). In all panels, $H$-band coronagraphic image by SUBARU \citep{Hioki+etal_2007} is shown in color scale. The white ellipse marks the 2.1 $\mu$m peak intensity of the disk with semi major axis of 3$\farcs$7 and inclination of 42$\degr$ determined by \citet{Close+etal_1998}.
Note that the ellipse center is 0$\farcs$7 off in Declination from the phase center.
The white square marks the peak at 2.1 $\mu$m and labeled as NIR 1 component by \citet{Hioki+etal_2007}. 
The grey triangle marks the SO south-east peak (see Sect. {\ref{sec:compact_co_so}}). 
}\label{fig:cont_pdb}
\end{center}
\end{figure*}
%
%

The CB and CS disks of UY Aur have been observed at various wavelengths.
At optical/NIR, a bright and patchy dusty CB ring inclined by $\sim$42$\degr$ and circumstellar materials connecting the CB ring and the CS disks were reported from observations with CFHT \citep{Close+etal_1998} and with SUBARU \citep[see Figure \ref{fig:cont_pdb}; ][]{Hioki+etal_2007}.
The brightest spot at NIR is on the south CB ring called NIR 1 \citep{Hioki+etal_2007}.
The CB disk itself is detected in $^{13}$CO 1-0 and 2-1 and yields an enclosed mass of about 1.2 M$_{\sun}$ \citep{Duvert+etal_1998}.
Besides the CB disk, there are extended ridges and hint of streamers seen in the measurements of the CO isotopologues.
At 3.6 cm the binary has also been detected using the VLA with a 2$\farcs$3 beam \citep{Contreras_2006}, which does not separate the two sources.
The spectral index of the UY Aur system was found to be surprisingly low, being $\sim$1.6 between 3.6 cm and 1.4 mm \citep{Contreras_2006}.

The photometric properties of the UY Aur binary are peculiar.
Until recently, UY Aur B has a high extinction at NIR while still exhibits a stellar spectrum (Herbst et al. 1995).
The $R$ band magnitude of UY Aur B changed by $\sim$ 5 mag within 50 yrs.
The $H$ band ($\lambda$ = 1.6 $\mu$m) magnitude of UY Aur B varied by 1.3 mag.
UY Aur A, which is relatively stable in $H$ band ($\triangle H$ = 0.4 mag) between 1996 and 2005 \citep{Close+etal_1998,Hioki+etal_2007},
exhibits a strong optical variability in $V$ and $B$ band \citep{Berdnikov+etal_2010}.

In this paper, we report our new observational results of the $^{13}$CO 2-1, C$^{18}$O 2-1, SO 5(6)-4(5) 
and $^{12}$CO 3-2 molecular transitions and of the thermal dust emissions at 1 mm and 3 mm.

\section{Observations}
\subsection{IRAM Plateau de Bure Interferometer}
The observations were carried out in 9 tracks on the IRAM Plateau de Bure Interferometer
(PdBI) from Dec.\,26, 2011 to Mar.\,16, 2012. At 1.4 mm (220 GHz), the observations were carried out in the B, C and D configurations, covering the baseline range of 15 m (11 k$\lambda$) to 456 m (338 k$\lambda$), in dual linear polarization.
The broad-band correlator provided an effective bandwidth of 7.6 GHz.
The total on source time is 44 hrs.
The narrow band correlator was set to cover the $^{13}$CO J=2-1 and C$^{18}$O J=2-1 lines with a
spectral resolution of 39 kHz (0.053 km/s).
The SO 5(6)-6(5) is detected in the broad-band correlator with a spectral resolution of 1.953 MHz (2.662 km/s).
At 3 mm (100 GHz), the observation were carried out in the A configuration, covering the baseline range of 110.5 m (37 k$\lambda$) to 666 m (223 k$\lambda$), with an on source time of 6 hrs.
All data were processed using the CLIC software in the GILDAS package.
Phase, amplitude, bandpass and flux calibration were performed following the standard procedure using 0400+258, 0459+252 for the amplitude and phase, 3C84 for the bandpass and MWC349 for the flux. 
To improve the image quality, we have applied phase-only self-calibration with a time interval of 45 second on the 1.4\,mm continuum, and transfer
the solution to the spectral line data.
For display, the channel maps of the $^{13}$CO 2-1 and C$^{18}$O 2-1 were
smoothed in velocity by 3-channel averaging.

The phase center is $\alpha$=04:51:47.388, $\delta$=30:47:13.1.
The effective angular resolution varies depending on the weighting scheme during inverse Fourier transformation to the image domain.
The highest angular resolution of 0$\farcs$55$\times$0$\farcs$34 for the 1.4\,mm continuum image is achieved using uniform weighting in order to resolve UY Aur A and B,
with an effective noise level (dynamic range limited) of 0.18 mJy/beam, or 1.8\,mK in brightness.
At 3\,mm, we reach an angular resolution of 0$\farcs$97$\times$0$\farcs$54 using robust weighting parameter 1, and a noise level of 0.04 mJy/beam, or 9.6 mK.
For $^{13}$CO, the angular resolution is 1$\farcs$27$\times$0$\farcs$92 obtained with robust weighting parameter 1, and the brightness
sensitivity is 17 mJy/beam, or 0.36 K with a channel spacing of 0.053 km\,s$^{-1}$ (the
spectral resolution being 1.6 times coarser due to channel shape).
The C$^{18}$O maps presented in Figure \ref{fig:chan_c18o} have been tapered to a
lower spectral angular resolution and the natural weighting scheme is used. The angular resolution is 1$\farcs$94$\times$1$\farcs$74, and the noise level is 12 mJy/beam, or 0.09 K.
The SO maps shown are obtained with robust weighting parameter 1 with an angular resolution of 1$\farcs$25$\times$0$\farcs$9 and a noise level of 1.2 mJy/beam. 



\subsection{the Submillimeter Array}
In addition to the PdBI observations,
we also obtained the $^{12}$CO 3-2 line data from the Submillimeter Array (SMA)\footnote{The
Submillimeter Array is a joint project between the Smithsonian
Astrophysical Observatory and the Academia Sinica Institute of
Astronomy and Astrophysics and is funded by the Smithsonian
Institution and the Academia Sinica.}.
The observations were carried out on Nov. 24, 2013 in the extended configuration.
The newly obtained data are combined with archival data taken in compact configuration on Dec. 13, 2010 in order to have better $uv$ coverage. The data were reduced using the MIR package following the standard calibration process.
The spectral resolution is 0.35 km/s.
The $^{12}$CO 3-2 maps are made with natural weighting, giving an angular resolution of 1$\farcs$12$\times$0$\farcs$98 at position angle (P.A.) 123$\degr$. The noise level at 1$\sigma$ is 0.17 Jy/beam, or 1.6 K.

\begin{deluxetable}{l | l l l}[!h]
\tablecaption{Flux densities toward UY Aur A and B} \tablewidth{0pt}
\tablehead{\colhead{$\lambda$} & \colhead{$S_{\rm A}$} &
\colhead{$S_{B}$} & \colhead{Reference} \\
\colhead{(mm)} & \colhead{(mJy)} & \colhead{(mJy)} & \colhead{}
}
\startdata
0.85 & 48.40$\pm$0.66 & 18.27$\pm$0.72 & ALMA, {\citet{Akeson+etal_2014}} \\
1.4 & 22.78$\pm$0.09 & 6.81$\pm$0.09 & PdBI, {This work} \\
3.4  & 5.4$\pm$0.1  & 1.7$\pm$0.1 & PdBI, {This work} \\
36 & 0.12$\pm$0.03   & 0.11$\pm$0.03 & VLA, {\citet{Contreras_2006}}\\
60 & $<$0.12 & $<$0.12 & VLA, {Data archive}
\enddata
\tablecomments{Columns are wavelengths ($\lambda$) in mm, flux density in mJy of UY Aur A ($S_{\rm A}$) and UY Aur B ($S_{\rm B}$) and the references.}
\label{tab:cont}
\end{deluxetable}

%
%
\begin{figure}[!ht]
\includegraphics[scale=0.35,angle=-90]{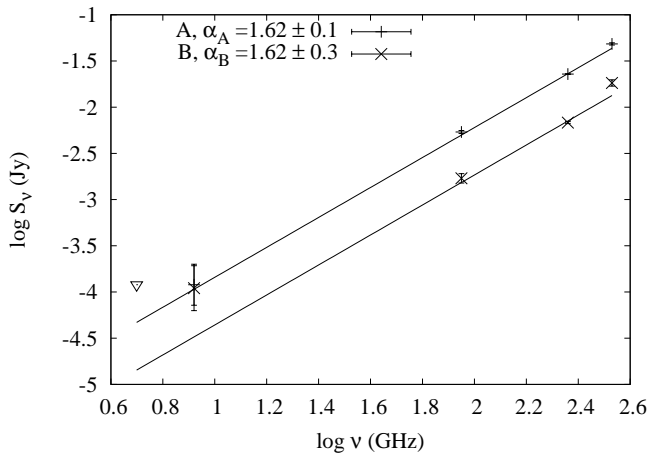}
\caption{Spectral energy distribution plot of UY Aur A and B. The triangle marks the 3$\sigma$ upper limit of $S_{\rm A}$ and $S_{\rm B}$ at 6 cm.}\label{fig:plot-sed}
\end{figure}

\begin{deluxetable*}{l l | l l | l}[!h]
\tablecaption{Continuum Best-fit Deconvolved size} \tablewidth{0pt}
\startdata
\hline \hline
\multicolumn{2}{c}{UY Aur A} & \multicolumn{2}{c}{UY Aur B} &   \\ \hline
Fitting Function & $\theta_{\rm maj}$, $\theta_{\rm min}$, P.A. & Fitting Function & $\theta_{\rm maj}$, $\theta_{\rm min}$, P.A. & $\lambda$ \\
   & ($\arcsec$),($\arcsec$),($\degr$) & & ($\arcsec$),($\arcsec$),($\degr$) & (mm) \\ \hline\hline
\multicolumn{5}{c}{Best-fit of the observed visibilities}  \\
Circular-Gauss & 0.21$\pm$0.01 & Circular-Gauss & 0.31$\pm$0.02 & 1.4 \\
Circular-Gauss & 0.21$\pm$0.01 & Elliptical-Gauss & 0.45$\pm$0.03, [0.2], 57$\pm$5 & 1.4 \\
Elliptical-Gauss & 0.24$\pm$0.01,0.19$\pm$0.01,110$\pm$7 & Elliptical-Gauss & 0.34$\pm$0.03, [0.2], 55$\pm$6 & 1.4 \\
Circular-Gauss & 0.08$\pm$0.03 & Circular-Gauss & 0.17$\pm$0.05 & 3.4  \\ \hline
\multicolumn{5}{c}{Deproject the global system with $i$=42$\degr$ and P.A. of the axis 45$\degr$}  \\
Circular-Gauss & 0.22$\pm$0.01 & Circular-Gauss & 0.38$\pm$0.02 & 1.4 \\
Circular-Gauss & 0.22$\pm$0.01 & Elliptical-Gauss & 0.46$\pm$0.03, [0.2], 18$\pm$7\tablenotemark{a}& 1.4
\enddata
\tablecomments{Columns are the fitting function ($f$), the de-convolved size ($\theta$ in arcsecond) of A and B, and the observed wavelengths ($\lambda$ in mm). 
The major and minor diameters and the P.A. are given.
The deprojection of the global system is done with an inclination of 42$\degr$ and P.A. of 135$\degr$.
The minor axis of the Elliptical-Gauss fit toward UY Aur B is fixed to 0$\farcs$2, shown as [0.2], in order to get a more robust fitting result.
}
\label{tab:cont_fit}
\tablenotetext{a}{The definition of P.A. here is in the rotated frame where the rotation axis of the circumbinary disk is on the y-axis. The P.A. will be 243$\degr$ in the sky frame.}
\end{deluxetable*}
\begin{deluxetable*}{l | l l l l l}[!h]
\tablecaption{Astrometry Parameters} \tablewidth{0pt}
\tablehead{ \colhead{Date} & \colhead{P.A.($\degr$)} &
\colhead{$q$(\arcsec)} & \colhead{P.A.$_{\rm deproj.}$ (\degr)} & \colhead{$q_{\rm deproj.}$ (\arcsec)}&  \colhead{Reference} }
\startdata
1944.25 & 212.2$\pm$0.4 & 0.82$\pm$0.1 & 215.4$\pm$0.4 & 1.09$\pm$0.13 & \citet{Joy1944}\\
1990 Nov 10 & 227.1$\pm$0.1 & 0.8750$\pm$0.0005 & 226.6$\pm$0.1 & 1.1771$\pm$0.0007 & \citet{Ghez+etal_1995} \\
1993 Dec 26 & 227.6$\pm$0.4 & 0.861$\pm$0.008 & 226.9$\pm$0.4 & 1.158$\pm$0.011 & \citet{Ghez+etal_1995} \\
1994 Sep 22 & 226.5$\pm$0.4 & 0.868$\pm$0.005 & 226.1$\pm$0.4 & 1.168$\pm$0.007 & \citet{Ghez+etal_1995} \\
1994 Oct 18 & 228.2$\pm$0.3 & 0.887$\pm$0.008 & 227.4$\pm$0.3 & 1.193$\pm$0.011 & \citet{Ghez+etal_1995} \\
1996 Oct 24 & 228$\pm$1 & 0.88$\pm$0.01 & 227$\pm$1 & 1.18$\pm$0.01 & \citet{Close+etal_1998}\\
1997 Dec 8 & 227.1$\pm$1.0 & 0.878$\pm$0.017 & 226.6$\pm$1.0& 1.181$\pm$0.023 & \citet{White2001} \\
1998.98    & 228.1$\pm$0.7 & 0.892$\pm$0.014 & 227.3$\pm$0.7 & 1.200$\pm$0.019 & \citet{Hartigan2003} \\
1999 Nov 16 & 227.6$\pm$0.4 & 0.88$\pm$ 0.01 & 226.9$\pm$0.4 & 1.18$\pm$0.01 & \citet{McCabe+etal_2006}\\
2000 Feb 22-23 & 228.82 $\pm$0.03 & 0.894$\pm$0.004 & 227.84$\pm0.03$ & 1.202$\pm$0.005 & \citet{Brandeker+etal_2003}\\
2002 Nov 23 & 229.13$\pm$0.04 & 0.886$\pm$0.001 & 228.07$\pm$0.04 & 1.191$\pm$0.001 & \citet{Hioki+etal_2007} \\
2005 Nov 9 & 231.36$\pm$0.04 & 0.890$\pm$0.001 & 229.74$\pm$0.04 & 1.194$\pm$0.001 & \citet{Hioki+etal_2007} \\
2011 Dec 30    & 231.22$\pm$0.32 & 0.94$\pm$0.013 & 229.63$\pm$0.32 & 1.26$\pm$0.013 & This work
\enddata
\tablecomments{Columns are the observation date (Date), the P.A. of the orientation between A and B (P.A.), the observed separation ($q$), the de-projected separation ($q_{\rm deproj.}$), the de-projected P.A. of the orientation (P.A.$_{\rm deproj.}$), and the reference.
The deprojection is calculated assuming i=42$\degr$ and axis P.A. of the CB disk of 45$\degr$. }\label{tab:separation}
\end{deluxetable*}

\section{Results}

\subsection{Continuum}
\label{sec:cont}
\subsubsection{compact continuum emission}
Continuum emission is detected at all three frequencies.
At 1.4\,mm, the emission is detected toward UY Aur A and B, which are clearly separated with the 0$\farcs$4 beam.
The 1.4\,mm and 3\,mm continuum overlaid on a SUBARU image is shown in Figure \ref{fig:cont_pdb}.
The coordinates and flux densities of A and B are determined by fitting the visibilities with two Gaussians.
The best-fit J2000 coordinates at the epoch of observations are
($\alpha$,$\delta$)=(04:51:47.392, 30:47:13.22) and ($\alpha$,$\delta$)=(04:51:47.335, 30:47:12.63), for the
continuum emission from UY Aur A and B respectively.
The flux densities of the UY Aur sources between 0.85\,mm and 6.0 cm are listed in Table \ref{tab:cont}.
We note that the flux density at 0.85 mm quoted in Table \ref{tab:cont} are from \citet{Akeson+etal_2014} due to a much higher angular resolution as compared to our SMA 0.85 mm measurement.
As discussed in \citet{Contreras_2006}, the 3.6 cm emission is consistent with the stellar location of the binary. At 6 cm, we obtained the archival data of the very large array (VLA) and analyzed them. There is no emission detected within the field of view, suggesting an upper limit of 0.12 mJy at 3$\sigma$.
We derive a spectral index $\alpha$, defined as $S(\nu) = S(\nu_0) (\nu/\nu_0)^{\alpha}$,
below 2  for both sources ($\approx 1.6$, see Figure \ref{fig:plot-sed}).

At 1.4\,mm, the emission around UY Aur A is well represented by a circular Gaussian
of deconvolved size $\sim$0$\farcs$2.  
The deconvolved size of $\sim$0$\farcs$2 probably reflects the residual seeing after self-calibration.
On the other hand, the emission around
UY Aur B appears more extended: an elliptical Gaussian fit yields a major axis ranging from 0$\farcs$34 to 0$\farcs$45, oriented along the AB line.
Using uniform disk models would yield larger sizes of the disks, i.e. 0$\farcs$34 and 0$\farcs$47 for A and B, respectively.
At 3\,mm, UY Aur B also appears slightly more extended (0$\farcs$17) than UY Aur A (0$\farcs$08),
although we have to fix the relative positions because of the more limited angular resolution.


Assuming the whole system is coplanar with the CB ring, we deproject the visibilities
of the 1.4\,mm continuum data with an inclination angle ($i$) of 42$\degr$ along P.A. 135$\degr$ in order to ease the comparison with previous studies. 
The emission from UY Aur A
again appears compatible with a circular Gaussian of size of 0$\farcs$22 (30 AU), which as mentioned
in the previous paragraph is most likely an upper limit due to seeing limitations.
The 1.4\,mm emission around UY Aur B is fitted by an elliptical Gaussian, with a best-fit major axis
of 0$\farcs$46 (65 AU) at P.A. 243$\degr\pm$7$\degr$, i.e. 18$\degr\pm$7$\degr$
away from the rotation axis of the system. 
The fitting results are listed in Table \ref{tab:cont_fit}.

We also derive a (deprojected) separation between the mm emissions around A and B of $1\farcs26\pm0\farcs013$,
at $\mathrm{P.A.} = 229\fdg6\pm0\fdg3$. This separation can be compared to the optical measurements
reported over the last decades (see Table \ref{tab:separation}), which are compatible with
a circular orbit of radius $\sim 1\farcs19$ \citep[167 AU][]{Hioki+etal_2007}. The mm separation is thus larger by 0$\farcs$07 at a $5 \sigma$ level. 
If we allow the orbit to be elliptical, by fitting a linear slope in
the deprojected separation, the best fit slope is 1.2$\pm$0.4 mas\,yr$^{-1}$,
and the 1.4\,mm position still deviates by $\sim 0\farcs05$
(4$\sigma$) from the extrapolated optical position.
We note, however, that the inclination angle is uncertain, a more detailed assessment of the UY Aur system
inclinations will be presented in Sect.\ref{sec:a_more_inclined_system}.

\subsubsection{Extended continuum emission}

At high spatial resolution, there is no visible emission outside the stellar positions, in particular
around the CB  disk/ring traced by NIR scattered light. However, the rms noise level, 0.18 mJy/beam
is relatively high at this resolution.
To trace the fainter extended continuum emission, we reconstructed the residual
image after subtracting the best-fit Gaussian of UY Aur A and B. We used natural weighting with a corresponding resolution 1$\farcs$9$\times$1$\farcs$69 in order to increase
the brightness sensitivity, reaching 0.11 mJy/beam or 0.86 mK.
The total flux in this image is 7.6 mJy. The total flux emanating from UY Aur system
at 1.4\,mm is thus about 38 mJy. This matches well with the 48$\pm$9 mJy measured
at 1.2\,mm  by \citet{Beckwith+etal_1990}, which gives $37 \pm 6$ after correction
for the mean spectral index of the emission.  We have thus recovered essentially
all the emission.

The resulting image is overlaid on the NIR emission in Figure \ref{fig:cont_pdb}.
The 1.4\, mm residual emission is brightest at 1$\farcs$5 away from the binary,
especially to the North and to the East, while the NIR emission is brighter South and West.
There is substantial mm emission near the NW circumstellar material identified
by \citet{Hioki+etal_2007} or the possible spiral arm identified by \citet{Close+etal_1998}, hereafter NW spiral, connecting from the CB ring to the inner disks.
Assuming a temperature of 20 K, and an opacity of gas and dust of 0.02 cm$^{2}$ g$^{-1}$,
the upper limit (at 3$\sigma$=2.6 mK) of the H$_2$ column density, $N_{\rm CB,gas}$,
is 4.3$\times$10$^{22}$ cm$^{-2}$ in the CB ring.  
The detected extended continuum peaks have column densities 2 to 4 times higher. The total continuum flux (7.6 mJy) recovered in the CB ring yields a total mass of $\sim 7\times10^{-4} \mathrm{M}_\odot$.

\begin{figure}[!h]
\includegraphics[scale=0.4]{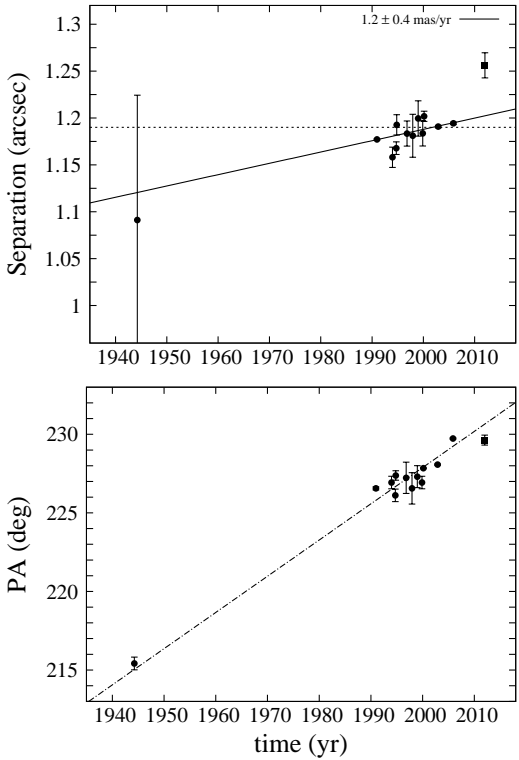}
\caption{Plots of the de-projected separation and position angle of UY Aur B with respect to UY Aur A at various epoch. The deprojection is done with an inclination of 42$\degr$ and P.A. of the disk rotation-axis 45$\degr$.
The dotted line marks a constant separation of 1$\farcs$19. 
The solid line marks the best-fit result of 1.2$\pm$0.4 mas yr$^{-1}$.
The dash-dot line marks the best-fit angular speed of 0.23$\degr\pm$0.03$\degr$ yr$^{-1}$.
The result from this work is shown in square.
}\label{fig:plots-sep-pa}
\end{figure}
%
%
\subsection{Gas}
\subsubsection{The extended gas traced with CO 2-1 isotopologues}
\label{sec:gas_detection}

\begin{figure*}[ht!]\center
\includegraphics[scale=0.65]{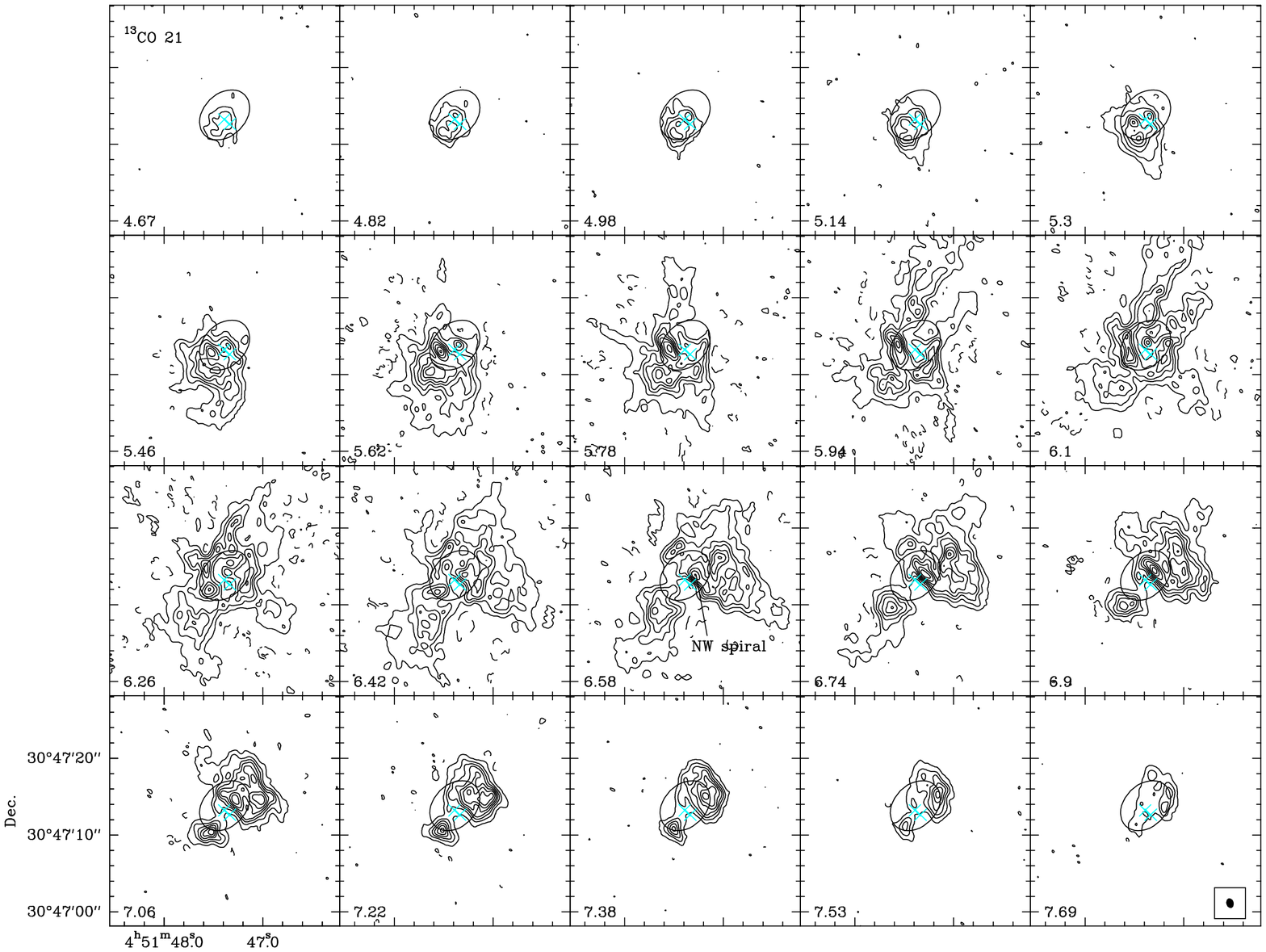}\\
\includegraphics[scale=0.65]{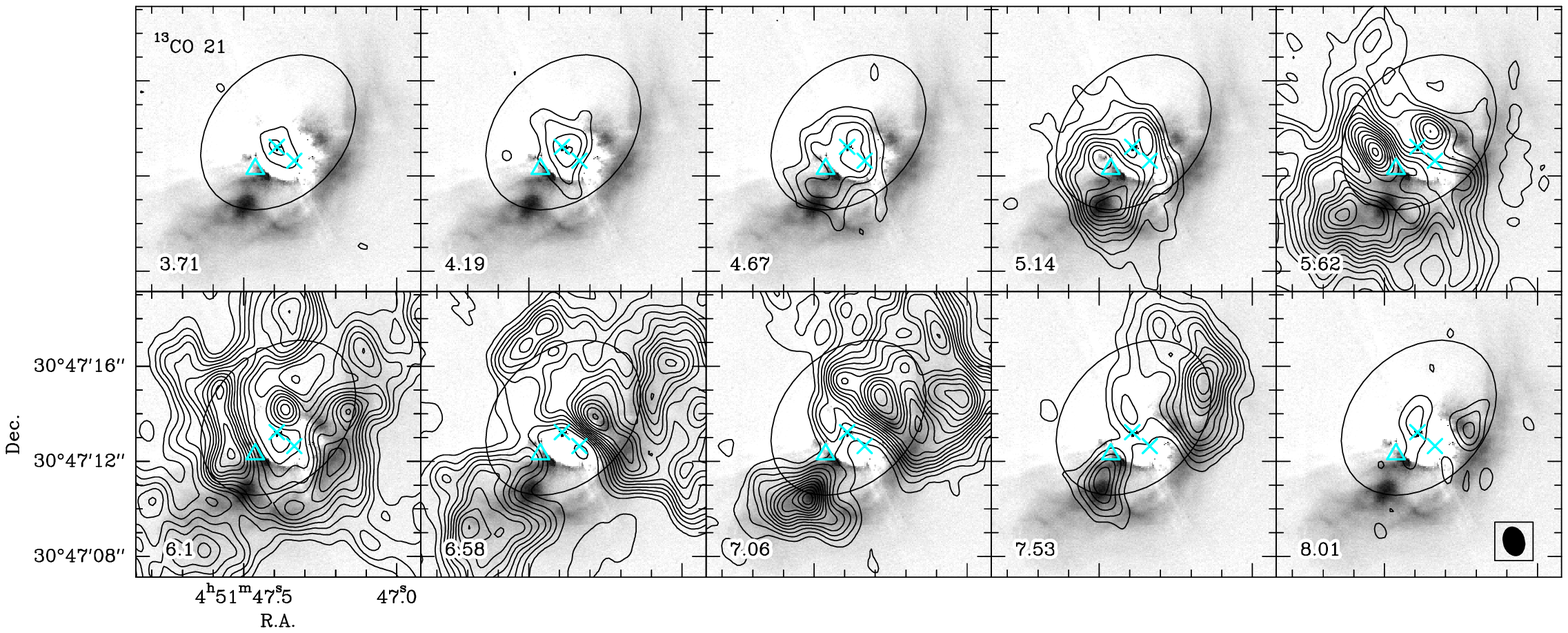}
\caption{Upper panels: Channel maps of $^{13}$CO 2-1. Contour starts at 29 mJy/beam (0.62 K; 3 $\sigma$) in steps of 6$\sigma$. Lower panels: zoom-in with higher velocity channels overlaying on the NIR image in grey scale. The rest of the symbols are the same as in Figure \ref{fig:cont_pdb}}\label{fig:chan_13co}
\end{figure*}

\begin{figure*}[ht!]\center
\includegraphics[scale=0.65]{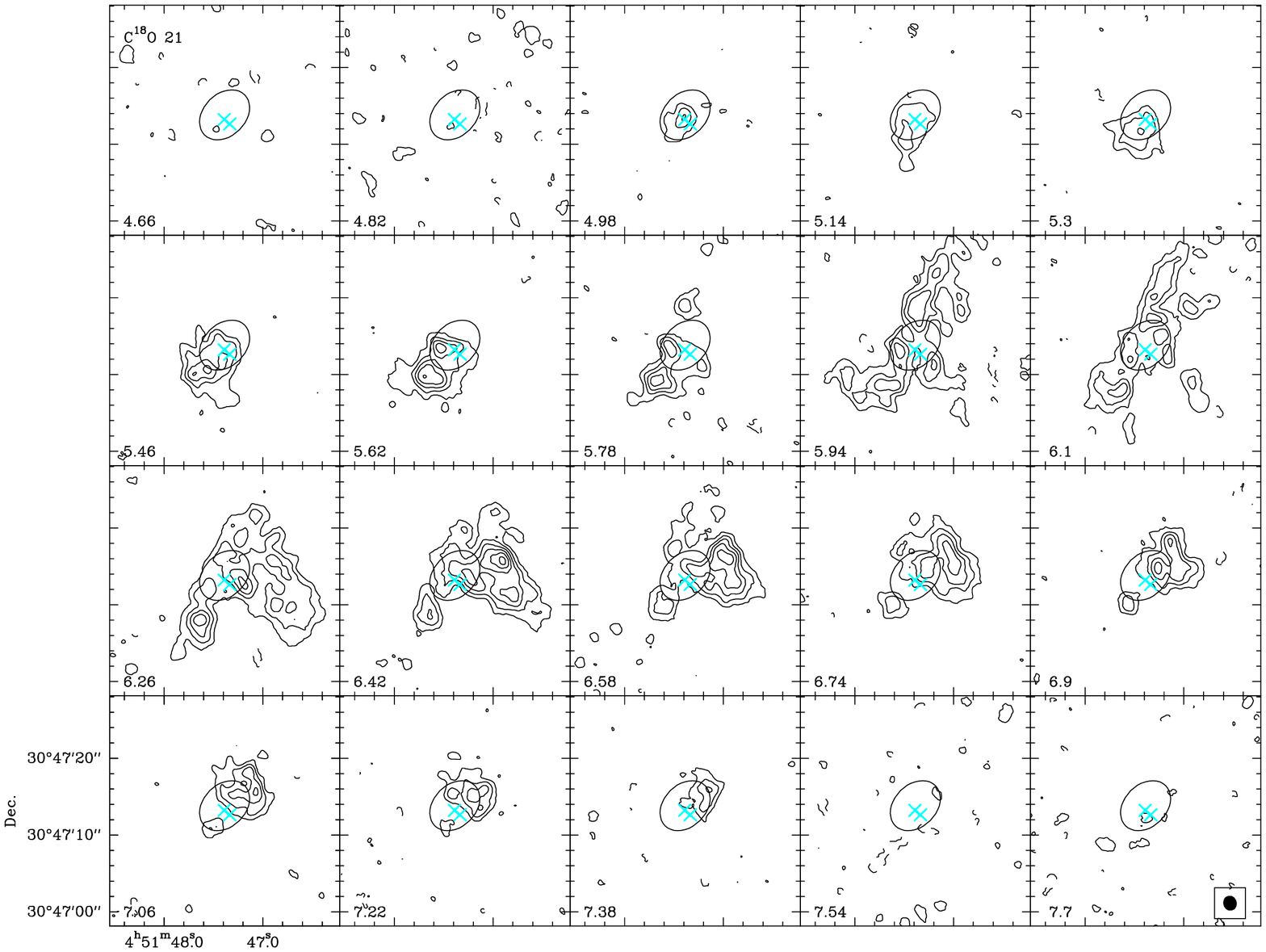}
\includegraphics[scale=0.65]{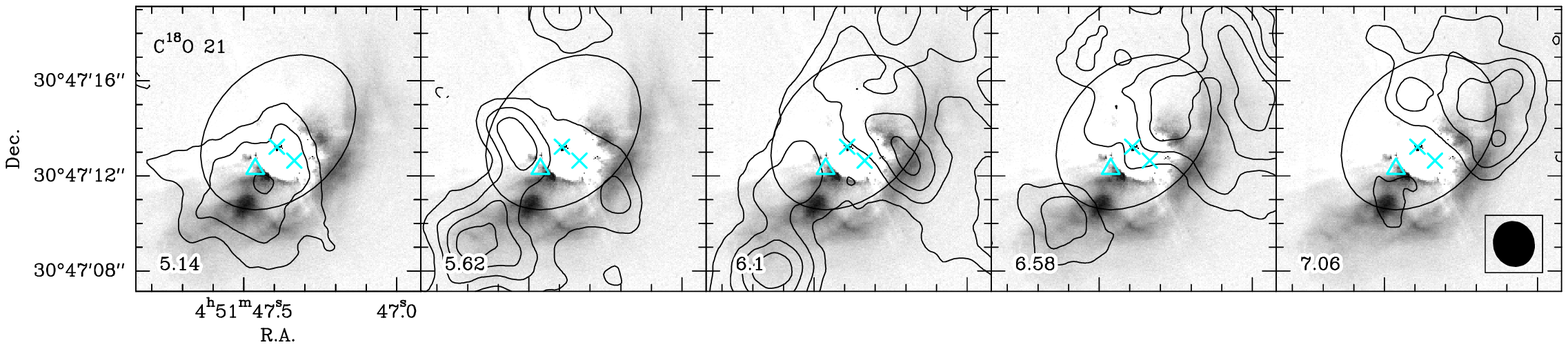}
\caption{The same as Figure \ref{fig:chan_13co} but for C$^{18}$O 2-1.  Contour levels are 20 mJy/beam (0.16 K; 3 $\sigma$). The rest of the symbols are the same as in Figure \ref{fig:cont_pdb}.}\label{fig:chan_c18o}
\end{figure*}

The $^{13}$CO 2-1 and C$^{18}$O 2-1 emissions are well detected and resolved by the new observations.
The channel maps of $^{13}$CO and C$^{18}$O are shown Figure \ref{fig:chan_13co} and Figure \ref{fig:chan_c18o}, respectively.
The total intensity maps of the four lines are presented in Figure \ref{fig:nir_m0}.
At low velocities, most of the extended emissions trace the CB disk/ring with peaks along the NW and SE directions, consistent with the proposed major axis of 135$\degr$ from the NIR image and with the previous images by \citet{Duvert+etal_1998}. 
However, the new millimeter data have twice higher angular
resolution, spectral resolution and brightness sensitivity which can reveal more details.

Several filamentary structures are detected.
There is a clear elongated structure with length of $\sim$10$\arcsec$ in the NW of the binary, especially at V$_{\rm LSR}$ of 5.94 and 6.1 km s$^{-1}$.
This northern filament connects to the CB disk near the systematic velocity, V$_{\rm sys}$, of 6.2 km s$^{-1}$, suggesting that it
is not just a random cloud along the line of sight, as initially found by \citet{Duvert+etal_1998}.
In addition, there are two new extensions, one from the CB ring at about 1$\arcsec$ NW of the UY Aur A (the NW spiral),
and another one with 6$\arcsec$ in length from the NIR 1 to the NE on the CB disk. 
The southern filament appears in the red-shifted velocities, and it connects to the NIR 1 peak.
Another extended gas structure at V$_{\rm LSR}$ of 6.42 to 6.74 km s$^{-1}$ is located to the west of the binary.
There is an emission deficit of $^{13}$CO and C$^{18}$O J=2-1 around the location of the binary, although there is some blue-shifted $^{13}$CO gas detected close to UY Aur A.
For the highest velocity components, we note that the blue-shifted velocity emission peaks close to the UY Aur A, while the red-shifted
velocity mainly comes from two regions: the first peak follows the CB ring in the NW where an extended
armlike structure is detected at NIR, and the second peak is around UY Aur A.
All the observed features traced with the $^{13}$CO and C$^{18}$O J=2-1 lines reveal that the CB ring is in close interaction with its
surrounding material, either with some remnant larger scale gas or with inner accretion disk(s).

Following the procedure described in \citet[Sect. 3.3.3]{Tang+etal_2012}, we estimate the gas column density, $N_{\rm H_2}$, of the CB ring from the $^{13}$CO 2-1 line.
The total emission of the CB ring except at NIR 1, which is probably contaminated by another cloud, is $\sim$0.49 Jy beam$^{-1}$ km s$^{-1}$ (or 10.4 K km s$^{-1}$).
Assuming an excitation temperature of $\sim 20$ K, an isotopic ratio $^{13}$C/$^{12}$C of 77 \citep[the value in local ISM]{Wilson+etal_1994}, and a CO/H$_{2}$ ratio of 10$^{-4}$.
we find  $N_{\rm H_2}$ is 4$\times$10$^{21}$ cm$^{-2}$, consistent with the upper limit of 4.3$\times$10$^{22}$ cm$^{-2}$ (Sect. \ref{sec:cont}) based on the 1.4\,mm continuum.
Thus, the detection of the CB ring dust continuum would require an improvement in sensitivity by about 1 order of magnitude (if the dust absorption coefficient per unit mass is correct),
down to 8$\mu$Jy beam$^{-1}$ at 1.4\,mm. The total gas mass (including both the gas mass of the ring and of the filaments) is $M_{\rm gas} \sim 2 \times 10^{-3}$M$_{\sun}$, based on the integrated area of the $^{13}$CO emission.

\subsubsection{Compact $^{12}$CO 3-2 and SO 5(6)-4(5) emissions}
\label{sec:compact_co_so}
\begin{figure*}[ht!]\vspace{1em}
\includegraphics[scale=0.7]{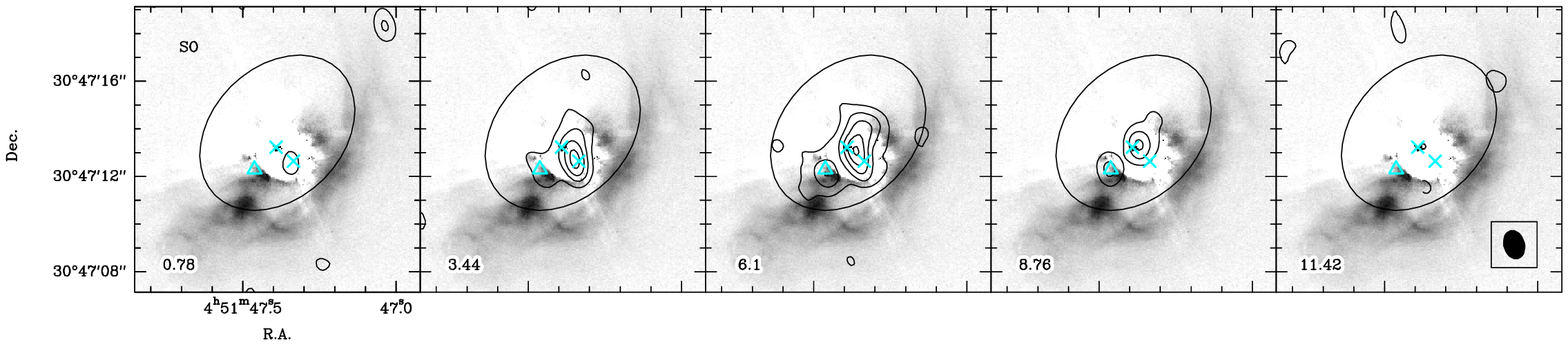}
\includegraphics[scale=0.7]{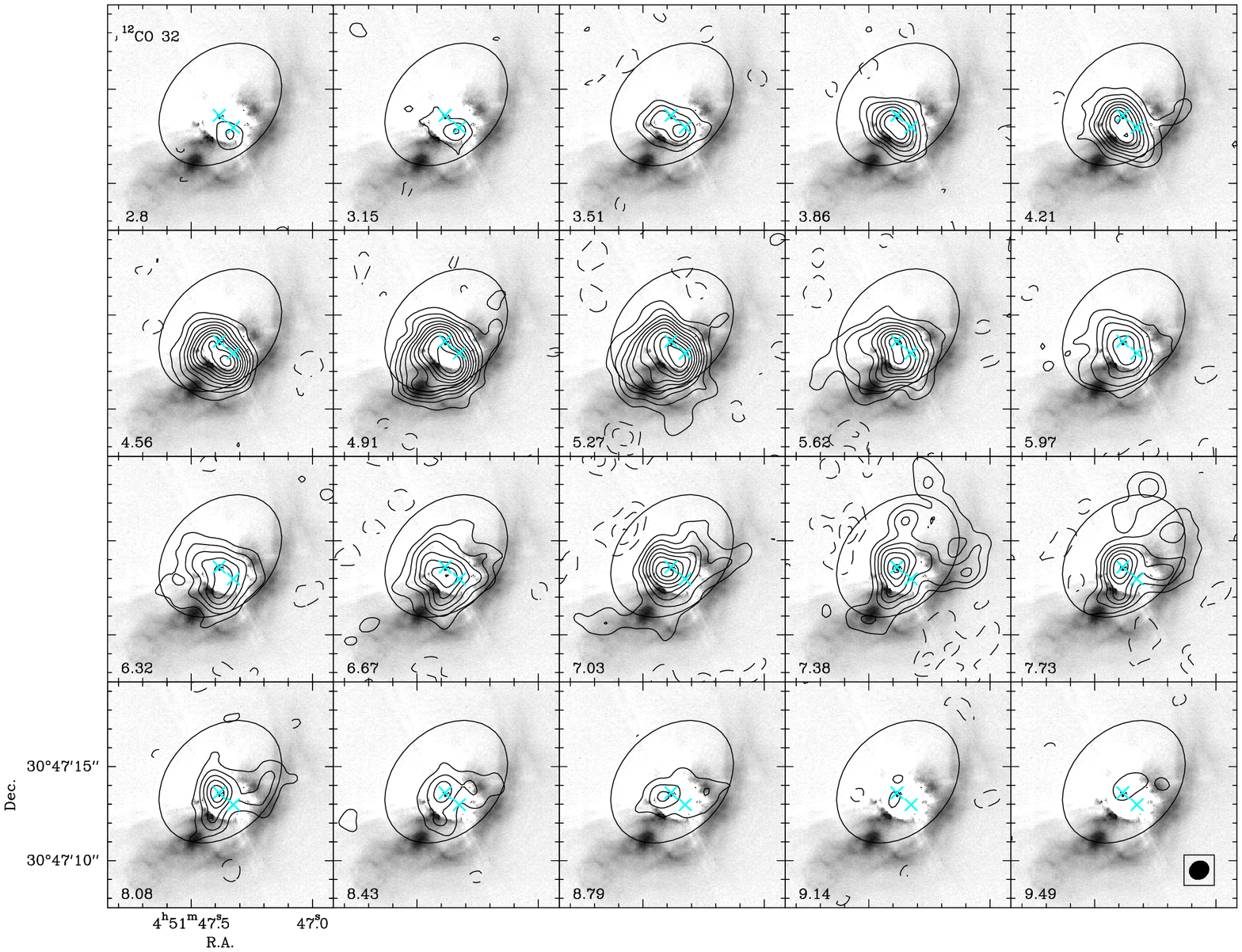}
\caption{Upper panels: SO channel maps overlaid on NIR image.
Contours start and step in 3$\sigma$ (1$\sigma$=1.2mJy per 1$\farcs$25$\times$0$\farcs$9 beam).
Lower panels: $^{12}$CO 3-2 channel maps. Contour starts at 5 K (3$\sigma$) and steps in 5 K.}\label{fig:chan_so_co32}
\end{figure*}

\begin{figure*}[!ht]
\includegraphics[scale=0.25]{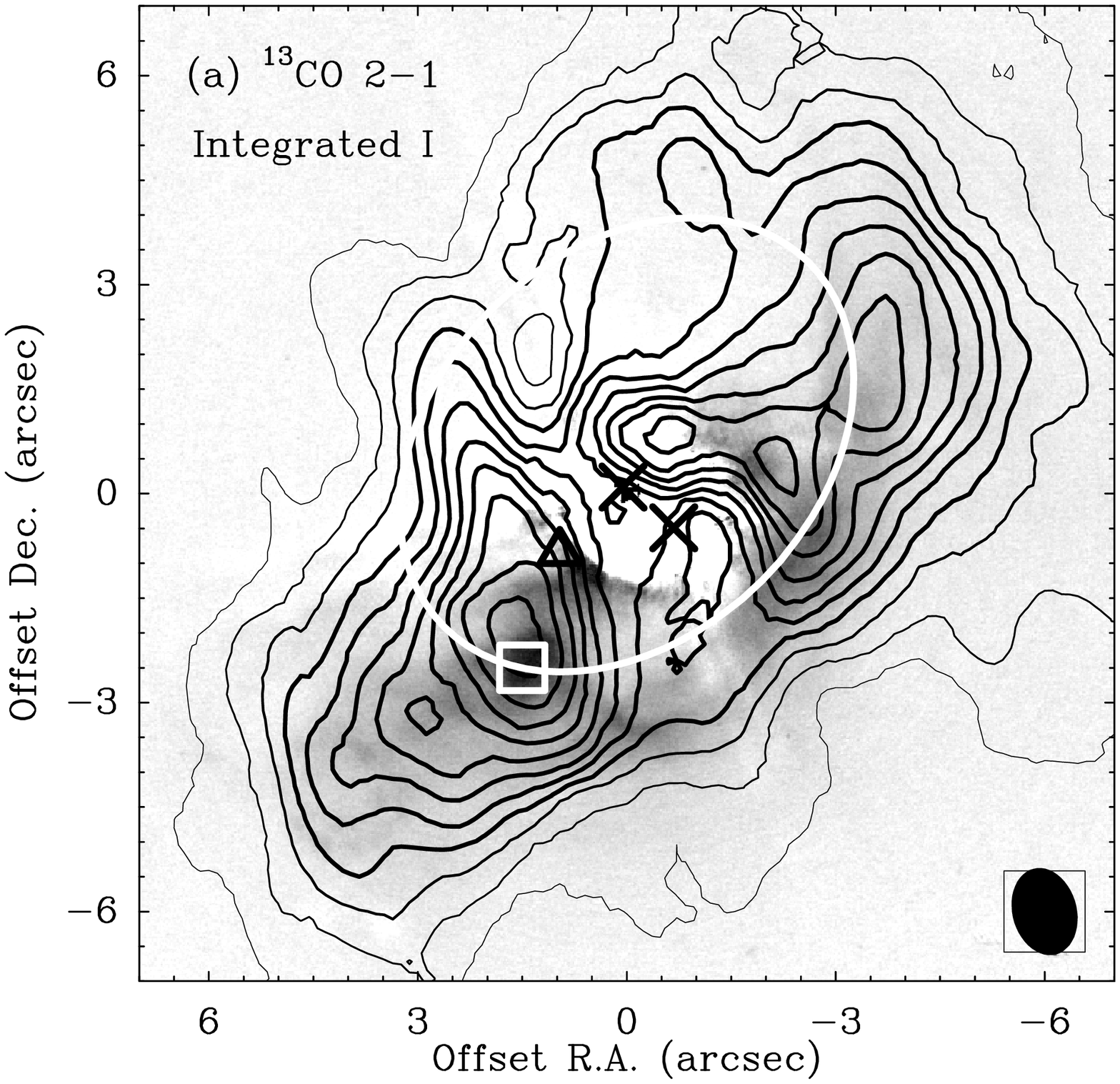}
\includegraphics[scale=0.25]{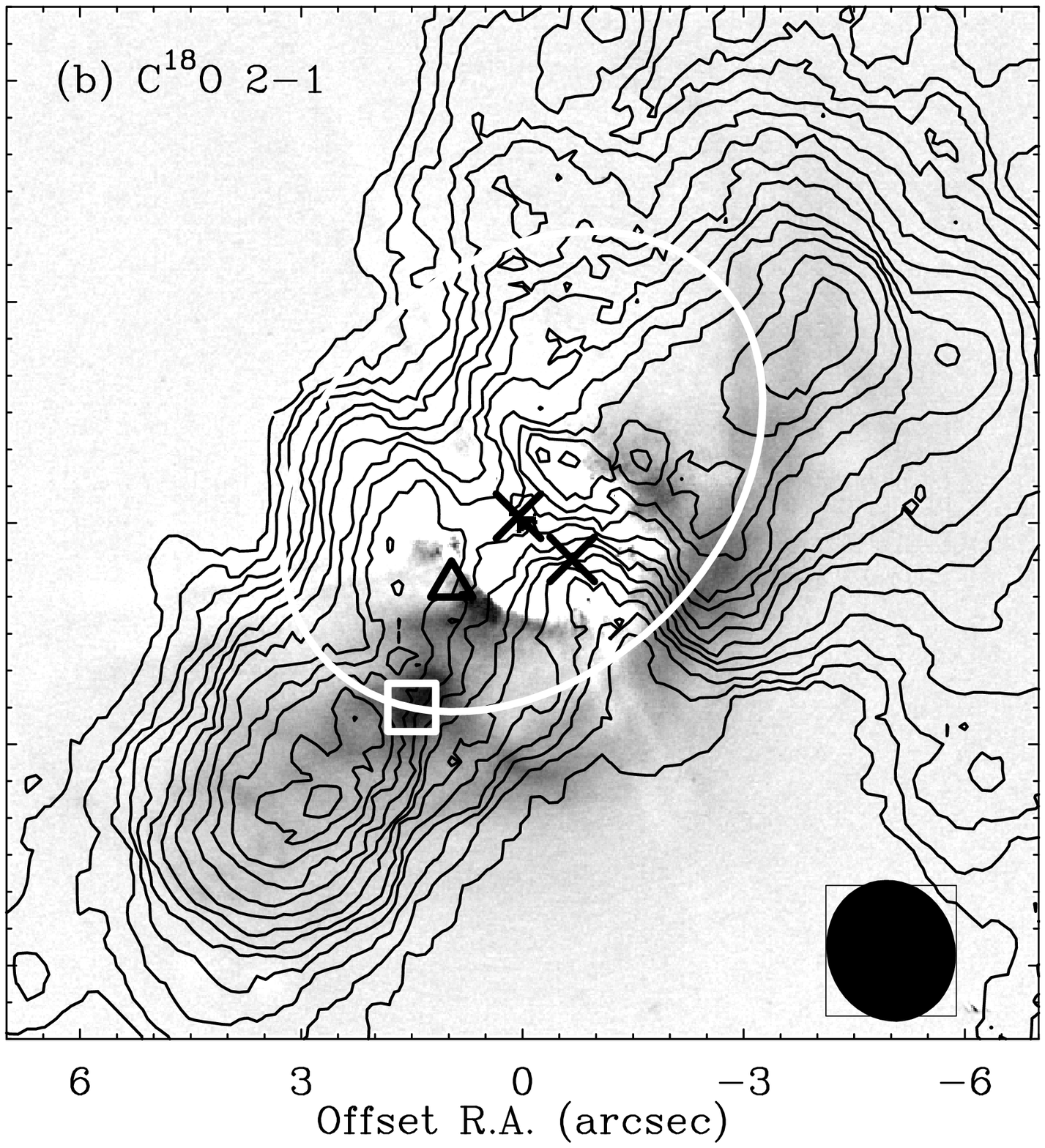}
\includegraphics[scale=0.25]{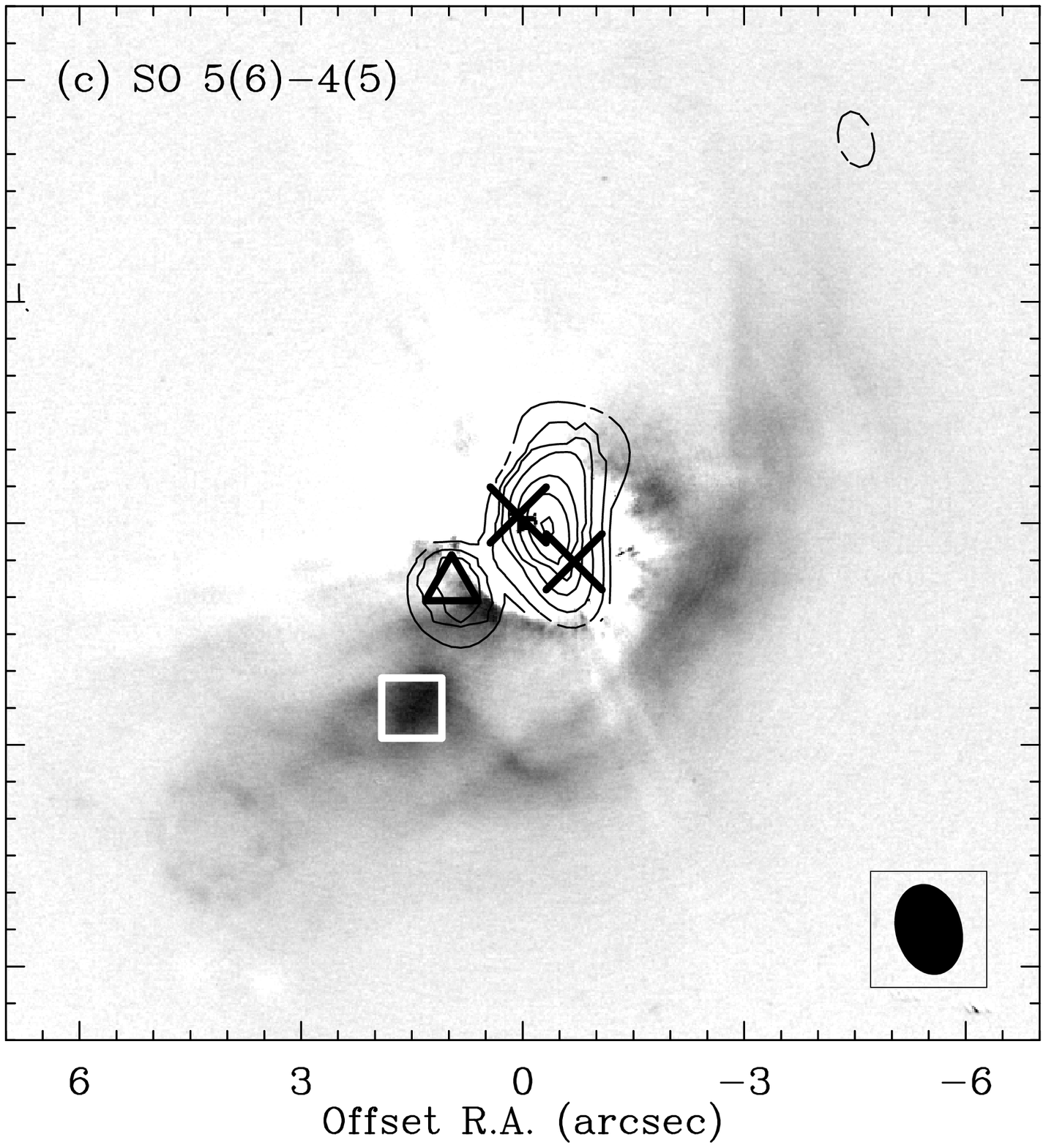}
\includegraphics[scale=0.25]{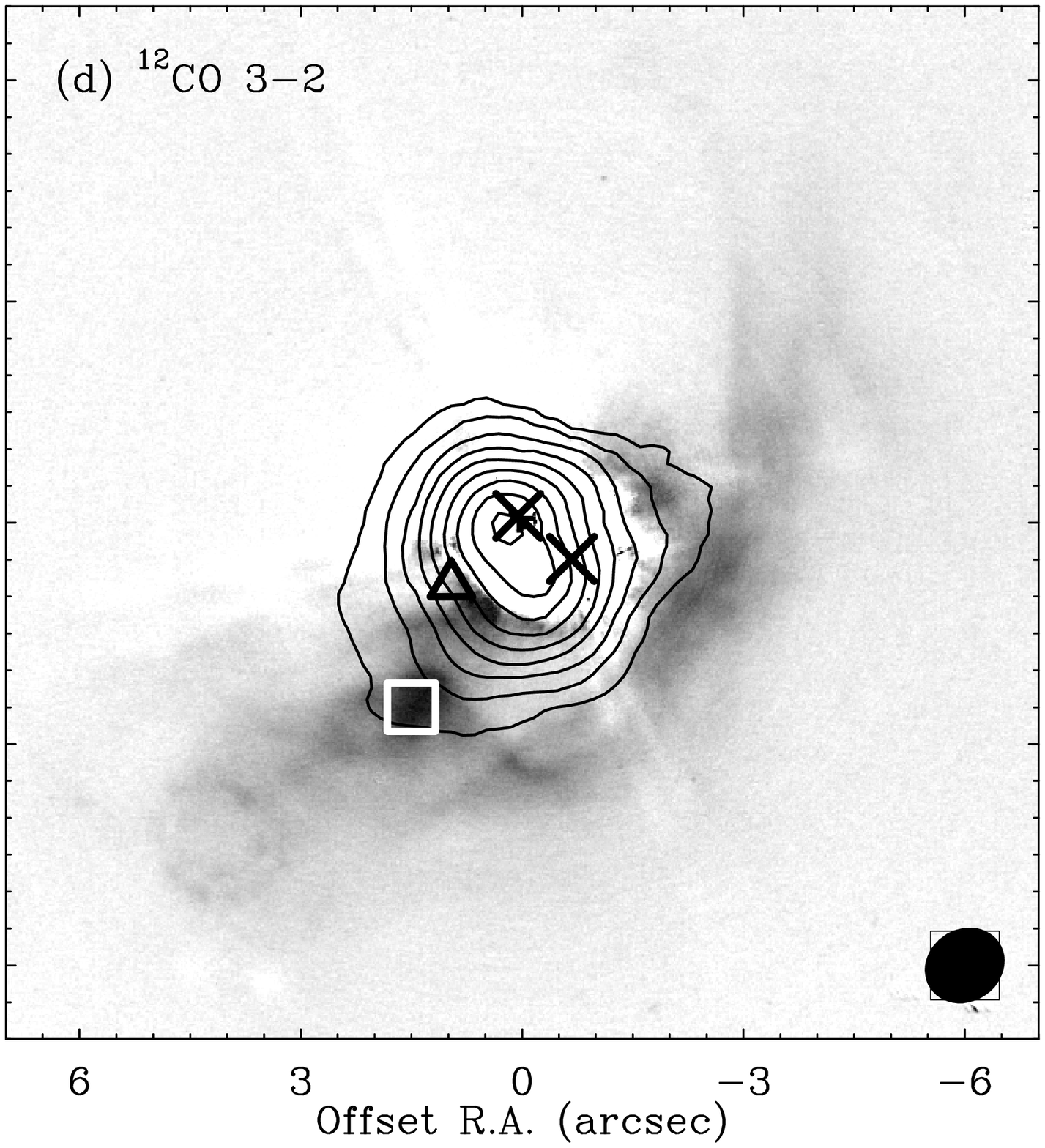}\\
\caption{Integrated intensity images of $^{13}$CO 2-1 (panel a), C$^{18}$O 2-1 (panel b), SO (panel c) and $^{12}$CO 3-2 (panel d) overlaid on the NIR image (grey scale).
In (a), the contour levels are 1, 2, 3, ...,11 $\times$ 0.05 Jy/beam km/s.
The beam size is 1$\farcs$27$\times$0$\farcs$92.
In (b), the contour levels are 1, 2, 3, ..., 12, 13 $\times$ 7 mJy/beam km/s.
The angular resolution is 1$\farcs$94$\times$1$\farcs$74.
In (c), the contour levels are 6, 12, 18, 24, 30, 36, 42 $\times$ 2.7 mJy/beam km/s (1$\sigma$).
The beam size is 1$\farcs$25$\times$0$\farcs$9.
In (d), the contour levels are 2, 4, 6, ..., 16, 18 $\times$ 1.1 Jy/beam km/s (1 $\sigma$).
The beam size is 1$\farcs$12$\times$0$\farcs$98 with P.A. of 123$\degr$.
}\label{fig:nir_m0}
\end{figure*}

The channel maps of $^{12}$CO 3-2 and SO are shown in Figure \ref{fig:chan_so_co32} and the total intensity maps are
presented in Figure \ref{fig:nir_m0}. The $^{12}$CO 3-2 emission appears very compact and peaks at UY Aur A,
either tracing the outflow(s), a disk or just warmer molecular gas. In contrast to CO, SO is often found in
outflow-driving and embedded sources \citep{Guilloteau+etal_2013}.
The SO emission in UY Aur is seen in between the binary with two
extensions: the main peak with extension toward the NW and the other peak, hereafter, SO SE peak, at 1$\farcs$5 SE to the main SO peak.
In summary, SO and CO 3-2 emissions are more compact than $^{13}$CO and C$^{18}$O 2-1 emission and peak closer to
the binary, likely tracing warmer gas than the $^{13}$CO and C$^{18}$O.

%
%
%
\begin{figure*}[!th]\center
\includegraphics[scale=0.75]{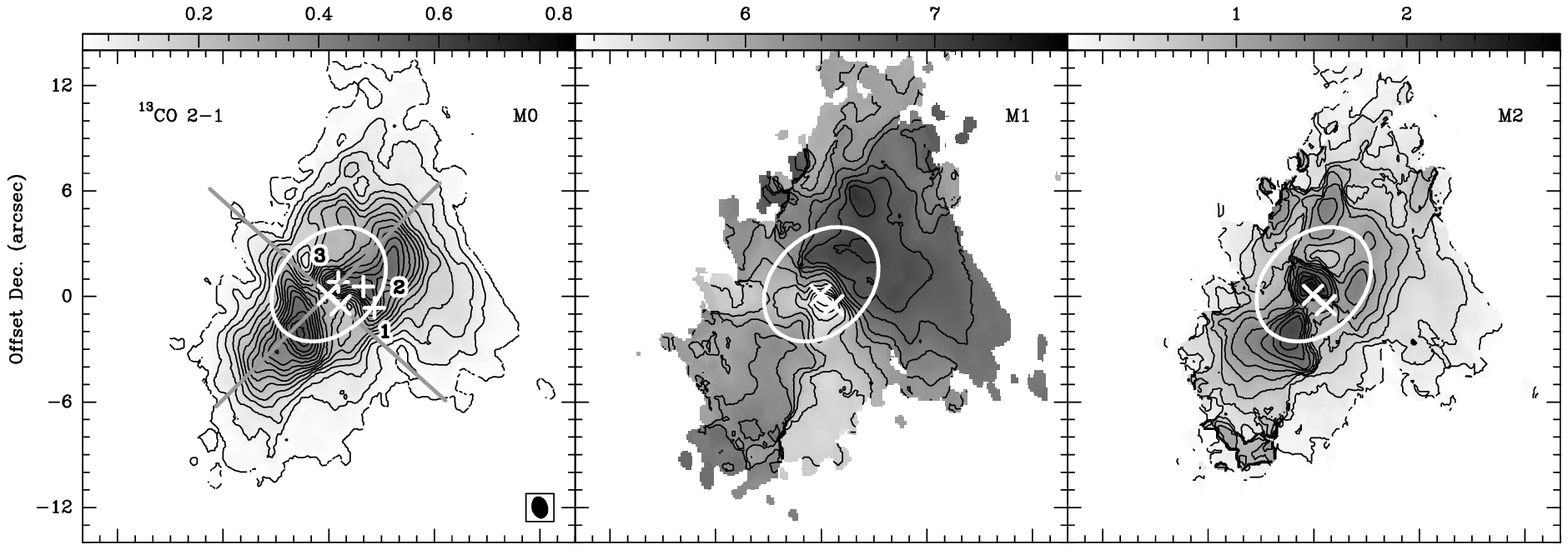}
\includegraphics[scale=0.75]{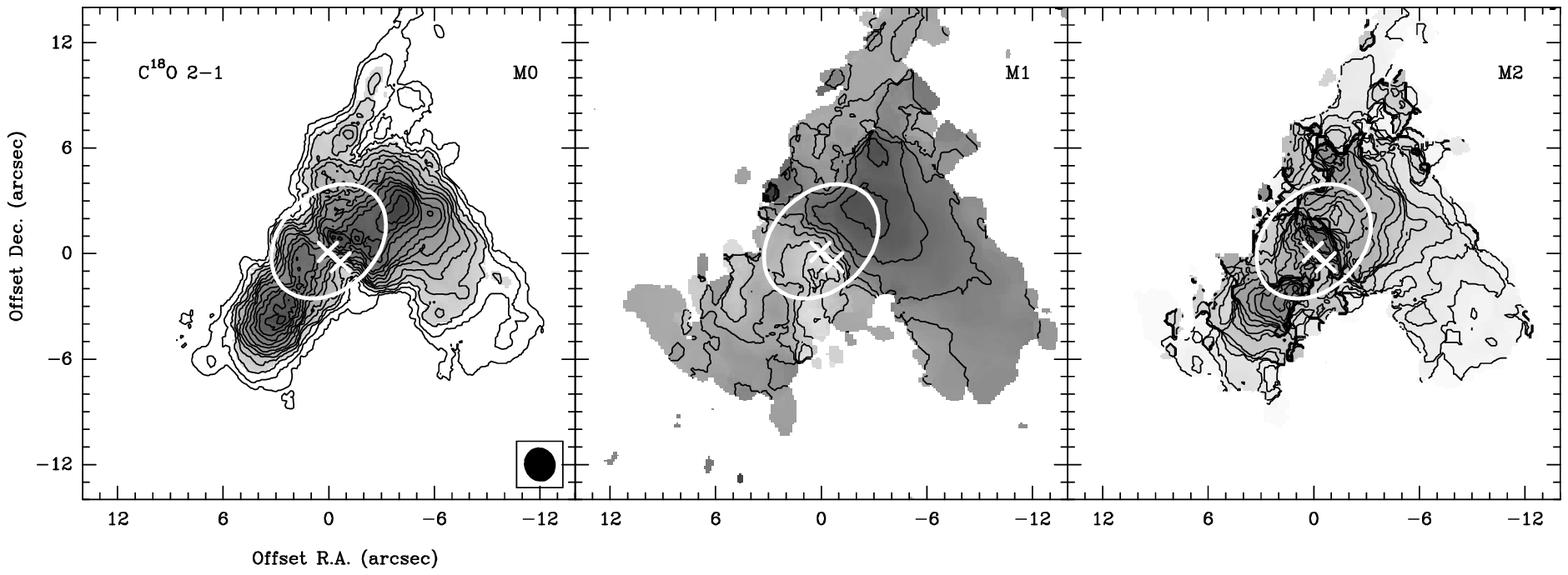}
\caption{Upper panels: Moment 0, 1 and 2 maps of $^{13}$CO.
The units of the wedge is in Jy beam$^{-1}$ km s$^{-1}$ for M0 and in km s$^{-1}$ for M1 and M2.
The three small "+" in (a) mark the positions NW Spi 1, 2 and 3 from west to east, where the spectra are taken.
Grey segments mark the position-velocity cut along P.A. of 135$\degr$ and along P.A. of 228$\degr$ (shown in Figure \ref{fig:pv}).
Lower panels: the same as upper ones but for C$^{18}$O.
Note that the intensity of the C$^{18}$O M0 map is scaled by a factor of 6.4.
}\label{fig:mnt}
\end{figure*}
%

\begin{figure*}[h!]
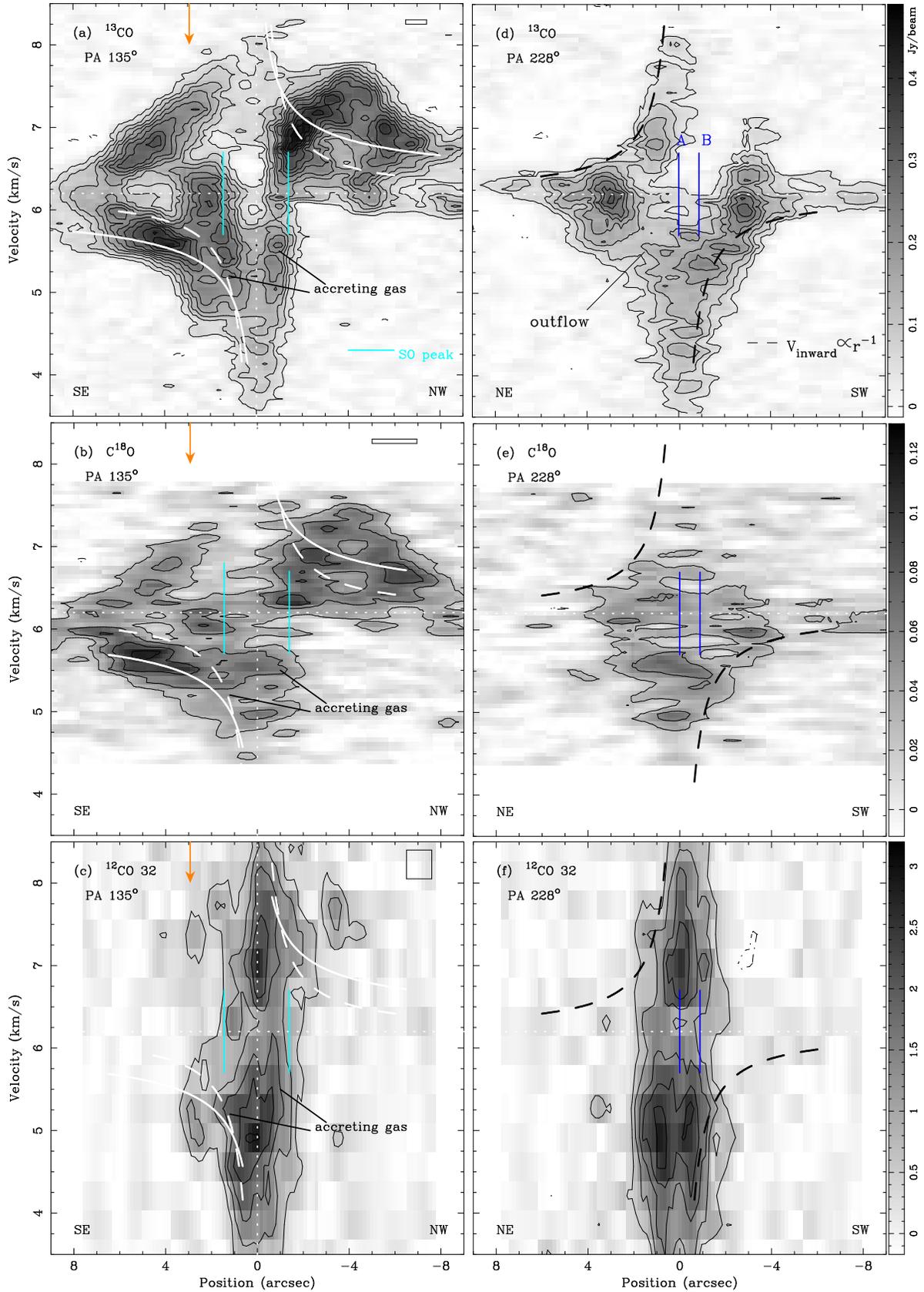

\begin{center}
\includegraphics[scale=0.45]{pv_13co.eps}
\includegraphics[scale=0.45]{pv_13co_pa228.eps}
\\
\includegraphics[scale=0.45]{pv_c18o.eps}
\includegraphics[scale=0.45]{pv_c18o_pa228.eps}
\\
\includegraphics[scale=0.45]{pv_12co32.eps}
\includegraphics[scale=0.45]{pv_12co32_pa228.eps}
\caption{Panel a, b and c: Velocity-Position plots along P.A. of 135$\degr$ centered on UY Aur A traced with  $^{13}$CO 2-1 (panel a), C$^{18}$O 2-1 (panel b) and $^{12}$CO 3-2 (panel c).
The solid curves mark the velocity of a Keplerian rotation with $i$=42$\degr$ and an enclosed mass of 1.2 M$_{\sun}$.
The dashed curves denote motions with conserved angular momentum (V$_{\rm rot}\propto$ r$^{-1}$).
The cyan segments mark the SO peaks. The orange arrow marks the NIR 1 location.
The resolutions are marked in the upper-right corner.
Panel d, e and f: the plots along the UY Aur A and B (P.A.=228$\degr$) traced with $^{13}$CO (panel d), with C$^{18}$O (panel e) and with $^{12}$CO 3-2 (panel f).
The blue segments mark the location of UY Aur A and B.
The black dashed curves denote the free-fall motion with $i$=42$\degr$ and  an enclosed mass of 1.2M$_{\sun}$.
In all panels, Vsys is 6.2 km/s (shown in dotted line).
}\label{fig:pv}
\end{center}
\end{figure*}

\subsubsection{A complex velocity pattern}
\label{sec:pv}

The velocity structure of the UY Aur system is very complex, as shown in both the moment maps
of $^{13}$CO and C$^{18}$O (Figure {\ref{fig:mnt}}) and the position-velocity (PV) diagrams (Figure \ref{fig:pv}).
The intensity-weighted velocity map (moment 1, M1, Figure \ref{fig:mnt}) of both $^{13}$CO and C$^{18}$O reveal that the main velocity gradient is along the major axis of the CB disk, suggesting that the velocity pattern is dominated by rotation at large scale.
Surprisingly, the velocity dispersion (moment 2, M2) is very large at the eastern edge of the CB ring and at UY Aur A. Although there is a general trend of red-shifted gas in the NW and blue-shifted gas in the SE,
the system contains a significant amount of gas deviating from simple Keplerian motions.

The kinematics is better analyzed using the position-velocity (PV) diagrams. In Figure \ref{fig:pv}a and \ref{fig:pv}b, the PV diagrams are
cut along the major axis of the disk (P.A. of 135$\degr$, shown as grey segments in Figure \ref{fig:mnt}).
In the blue-shifted part, there are three gas components with velocity gradient: the outer most one is in Keplerian rotation
and is probably tracing the extended CB disk, and the other two are close to the SO extension peaks.
We note that there is about comparable emission strength between the extended CB disk and the ones associated with the SO extension peaks. 
Interestingly the PV diagrams along the binary (AB line) of 228$\degr$ (Figure \ref{fig:pv}d, \ref{fig:pv}e), which
is almost perpendicular to the major axis of the disk, also show some
high velocity streamers in between UY Aur A and B. Such a velocity pattern is characteristics of infalling material
as illustrated by the black dashed curve ($V_{\rm infall} \propto r^{-1}$) in panel (d) and (e) and likely
traces gas coming from the CB ring which is accreted onto the individual disk components.
There is also an outflow component detected at 1 to 2$\arcsec$ in the blue shifted part near UY Aur A (Figure \ref{fig:pv}d,e).
The PV plots of $^{12}$CO 3-2 along P.A. of 135$\degr$ and along P.A. 228$\degr$ are shown in Figure \ref{fig:pv}c and \ref{fig:pv}f, respectively.
Contrary to CO 2-1 gas, the $^{12}$CO 3-2 only partly trace the CB ring (panel c, P.A. of 135$\degr$) and the infalling
material (panel f, P.A. of 228$\degr$). It essentially traces warmer gas located near the stars and which is not seen in
$^{13}$CO and C$^{18}$O 2-1 maps which characterize colder material.

We further extract the spectra of $^{13}$CO 2-1, C$^{18}$O 2-1 and $^{12}$CO 3-2 at UY Aur A and B, at NIR 1, at the SO SE peak and along the NW spiral (Figure \ref{fig:spec}).
The locations where the spectra are taken along the NW spiral are marked in Figure \ref{fig:mnt}.
In general, the $^{13}$CO and C$^{18}$O show similar behavior, as was seen in Sec. \ref{sec:gas_detection} and \ref{sec:pv}.
The $^{12}$CO 3-2 spectra have dramatically different shapes from $^{13}$CO and C$^{18}$O along the NW Spi1 and NW Spi2, and at NIR 1.
There is no good reference velocity to compare with the obtained spectra, because the dominant force is highly dependent on the distances to UY Aur A or B.
We note, however, at the NW spiral, there are clearly high velocity wings seen in $^{13}$CO in the red-shifted end, especially in Spi1 and Spi2.

\section{Discussion}\label{sec:discussion}

\subsection{The extended CB ring}

The CO emission is clearly dominated by the extended CB ring, but there is only a marginal detection
of a dusty counterpart (Figure \ref{fig:cont_pdb}). This suggests that the amount of material in
the CB ring is low, as estimated from both the $^{13}$CO 2-1 integrated emission and the 1.4 mm residual dust emission.
The total mass (gas+dust) derived from the extended $^{13}$CO emission around UY Aur (hence including the streamers) is of the order of $M_{\rm gas} \sim 2 \times 10^{-3}$ M$_{\sun}$. 
The mass derived from the residual extended flux emission is $\sim 7\times10^{-4}~\mathrm{M}_\odot$.
Hence the masse of the ring, is of the order of $10^{-3}$ M$_{\sun}$. This is two magnitudes lower than that of the ring orbiting GG Tau A \cite{dutrey1994}.
Contrary to the GG Tau ring \citep{dutrey1994}, where the inner edge of the CB ring is well defined, the inner edge of UY Aur is at $\simeq 540$\,AU, in agreement with \citet{Duvert+etal_1998}, is not well defined. 
The contrast between the CB ring and cavity between the CB ring and CS disks is moderate, as there is substantial gas emission in the cavity. 
There is about comparable amount of emission inside the cavity as on the CB ring (see Sect. \ref{sec:gas_detection} and \ref{sec:pv}),
making the determination of ring size, its sharpness and its inclination more difficult.

\begin{figure*}[ht!]
\includegraphics[scale=0.6]{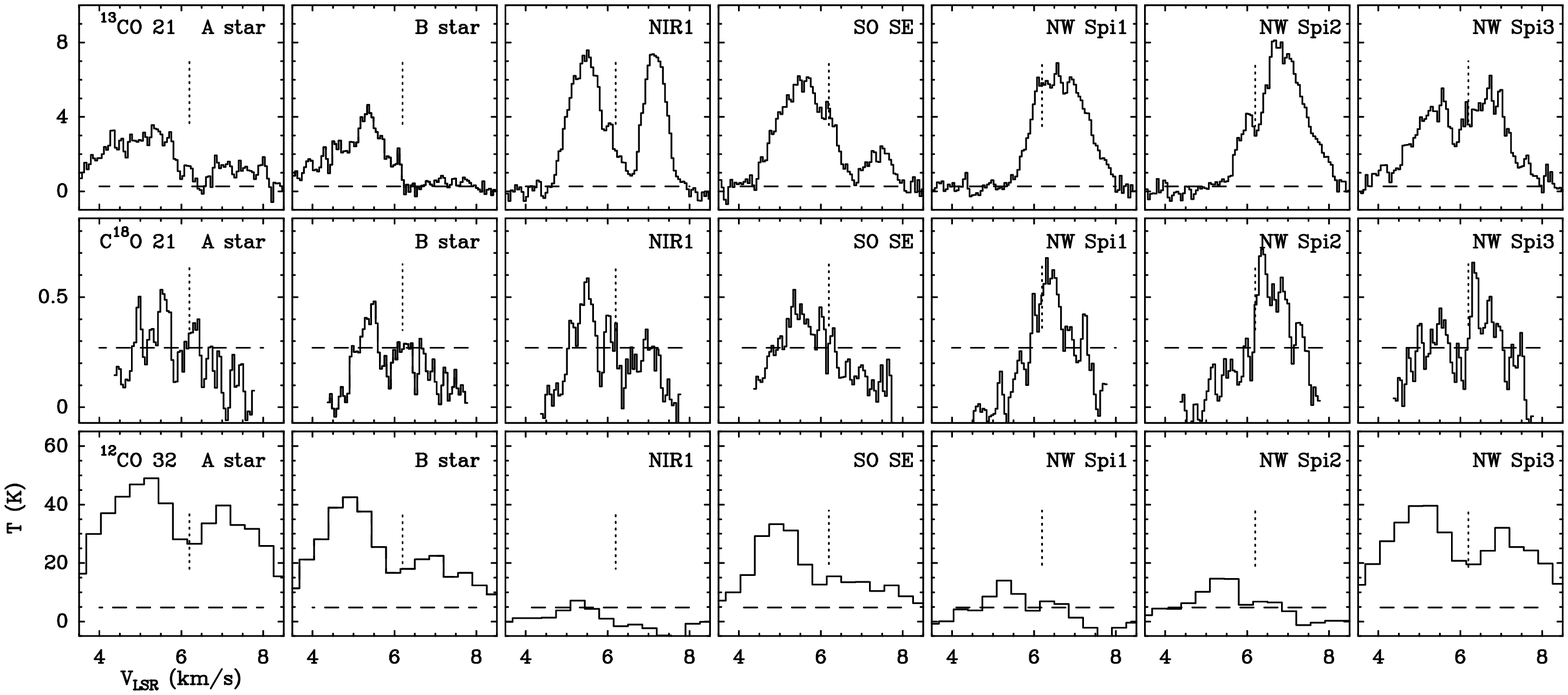} 
\caption{Spectra of $^{13}$CO 2-1, C$^{18}$O 2-1 and $^{12}$CO 3-2 toward UY Aur A, UY Aur B, the NIR 1 peak, the SO south-east peak, and three spots of the NW Spiral marked as "+" in Figure \ref{fig:nir_m0}.
The vertical dotted line marks the V$_{\rm sys}$ of 6.2 km/s. 
The dashed lines mark the 3$\sigma$ level.}\label{fig:spec}
\end{figure*}

\citet{Hioki+etal_2007} concluded that scattering anisotropy was insufficient to explain
the strong contrast between the NW and SE parts of the CB ring, and that the far side of the ring
was obscured by additional material. The anti-correlation between the NIR scattered light
and the detected mm continuum emission provides direct support for this selective extinction
hypothesis. The extended $^{13}$CO emission (see Figure \ref{fig:mnt}) toward the East of the binary further supports this interpretation.
We note that the kinematics is peculiar at NIR 1 peak.
At the location of NIR 1 which is at the near side, there is an additional $^{13}$CO emission peak at red-shifted velocity V$_{\rm LSR}$ of 7.16 km s$^{-1}$ besides the blue-shifted Keplerian rotation (Figure \ref{fig:spec}). This extended component has a size of $\sim$5$\arcsec$ with a projected velocity gradient of 1.4 km s$^{-1}$ (Figure \ref{fig:pv}). 
This cloud is likely gravitationally bound to the binary system since its velocity difference to the Keplerian speed at this location is only $\simeq 0.6-1.4$ km s$^{-1}$.
Like in the case of AB Auriga \citep{Tang+etal_2012}, such a feature may be a foreground CO cloud captured by the UY Aur system and falling down to the CB material, where the emission appears red-shifted due to projection.
This in falling gas at NIR 1 provides an additional explanation of the contrast of the CB ring at NIR.

%

\subsection{Evidences for streamers}

Several extended features suggest the existence of accretion material between the CB disk and the CS disks.
At NIR, a hint of possible spiral arm is identified in $J$ band by \citep{Close+etal_1998}.  
In $H$ band \citep[Figure \ref{fig:cont_pdb};][]{Hioki+etal_2007} there are two features identified at the base of the inner edge of the ring, and one of which corresponds to the possible spiral arm in $J$ band. 
An $H_{2}$ 2.12 $\mu$m arc of $\sim$ 0$\farcs$5 in extension is also detected in the south
of the CS disk of UY Aur A toward the east by \cite{Beck+inprep}. Such an extension cannot be explained by the outflow since the
bipolar outflow has a P.A. of 40$\degr$ \citep{Hirth1997}.
Although the excitation conditions are not yet determined, the detected $H_{2}$ arc may be due to accretion shocks onto
the CS disk surrounding UY Aur A, as it is observed in the case of GG Tau Aa \citep{Beck+2012}.
In addition to the $H_{2}$ arc, diffuse $H_{2}$ emission is observed around both stars.

Our new detections of the SO, $^{12}$CO, $^{13}$CO and C$^{18}$O gas surrounding the binary further support the existence of streamers.
In the integrated intensity images, the NW accreting material is clearly resolved in $^{13}$CO and C$^{18}$O (Figure \ref{fig:nir_m0}).
Both in $^{13}$CO and C$^{18}$O 2-1, the cavity does not appear free of gas. At the inner edge of the CB disk, the emission  is asymmetric (Figure \ref{fig:mnt}). 
Moreover, at the location of the SO extention peaks, there is some CO gas moving with a velocity
gradient of 1 to 2 km s$^{-1}$ traced in $^{13}$CO and C$^{18}$O (see the PV plots in Figure  \ref{fig:pv}a,\ref{fig:pv}b; spectra in Figure \ref{fig:spec}). 
In the two lines, confusion along the line of sight due to the NIR 1 component also precludes a good identification of the the SE spiral.
Contrary to $^{13}$CO and C$^{18}$O 2-1, the $^{12}$CO 3-2 emission is more compact and confined within the gap of the CB disk (Figure \ref{fig:nir_m0}).
Such a difference, likely due to a temperature effect in the ring, also suggests that there are gradients of excitation
conditions inside the ring. 

The SO emission peaks around the binary with east and west extensions along the axis perpendicular to the binary.
We also note that the detected SO extensions are closer to the primary UY Aur A instead of the secondary.
SO is known to be a shock tracer. Although we cannot reject that some SO emission may come from the outflow, the extended SO emission may come from a shock point at
the interface between the inner edge of the CB disk and the inward accreting material, as predicted by e.g.
\citet{Hanawa+etal_2010}, where the gas inflows from the CB disk through L2. Based on simulations and depending on the
binary orbit, its eccentricity and mass ratio, this gas is further captured mainly by the primary \citep[e.g.][]{Ochi+etal_2005}
or by the secondary \citep{Bate1997}.
Future ALMA observations with multi-J transitions of the CO isotopologues and the SO lines at higher angular resolution would provide
a clearer picture of the streamers and their links with both the CB and CS material.

\subsection{Material surrounding the individual stars}

When UY Aur was identified as a binary by \citet{Joy1944}, the two stars were optically visible.
UY Aur B has been once classified as an infrared companion \citep{Tessier+etal_1994}.
Its magnitude changed on a long period by several magnitudes since 1944 \citep{Berdnikov+etal_2010}.
Interestingly, several observational evidences also indicate that the accretion process around
the binary varies significantly within one decade. \citet{Skemer+etal_2010} report
that both UY Aur A and B varied at mid-infrared between 1999 to 2009, and that the dust grains
surrounding UY Aur A are extremely pristine and ISM-like based on the silicate spectrum.
\citet{Skemer+etal_2010} interpret such dust properties as an indirect proof of replenishment
by a less evolved CB dust through spiral streamers.
Similarly, around UY Aur B, \citet{Herbst+etal_1995} argued that such changes in magnitude cannot be caused
by spherically distributed material, but may be caused by inhomogeneous accretion from
a flattened structure.
Moreover, \citet{Herbst+etal_1995} reported the existence of an outflow while
\citet{Pyo+etal_2014} revealed that both UY Aur A and B drive optical jets
along an axis at P.A. of $\sim$40$\degr$, in agreement with the outflow axis.
Finding a scenario which reconciles all particularities of the binary system seems to be difficult.
We discuss in this section some facts, their implications on the system and some possible ways to reconcile all
the observed properties of the source.

\subsubsection{UY Aur A}
\label{sub:discussion:uyaura}

We first note that the low spectral index of the mm emission, $\sim 1.6$, can only
be explained either by optically thick dust emission at low temperatures (10 K, with an
upper limit of 20 K given the uncertainties), or by a superposition of warmer dust and free-free emission at longer wavelengths. 
As the emission originates from distances smaller than
20 AU from the star, the required temperature appears unrealistically low.
Free-free emission from an ionized jet is expected to have a spectral index of 0.6, while dust at
25 K would have an index of 1.9. The kink in the spectral-energy-distribution indicated by the 0.85\,mm flux measurement
made with ALMA by \citet{Akeson+Jensen_2014} further suggests a contamination by free-free
emission, which becomes negligible only above 300 GHz. In this scenario,
we may attribute 18 mJy to the dust emission at 220 GHz, which requires a
minimum radius of 12 AU using the typical compact disk properties
as described in \citet{Pietu+etal_2014}.

\citet{Berdnikov+etal_2010} discussed the $B$ and $V$ band light curves on a timescale of 100 years as the Infrared variations of UY Aur A and B. They identified a periodicity of 16.3 years in
the variations of UY Auriga A, which they interpreted as a modulation of the accretion
rate induced by a companion located at 6 AU from the star.
Our detection of mm continuum emission restricts the possibilities for such a companion.
The invoked companion should not be too massive, in order
to avoid opening a substantial gap in this disk. Placing a firm upper limit
on the body mass is however premature, as the width and depth of tidal gaps
also depend on the disk viscosity \citep{crida+2006}.

\subsubsection{UY Aur B as a binary ?}
\label{sec:discussion:uyaurb}
Besides the low spectral index, the mm emission around UY Aur B has two further unusual
characteristics: 1) it is elongated along the overall UY Aur system rotation axis, and 2)
it is offset by 0$\farcs$07 (deprojected, 10 AU) from the optical position of UY Aur B along the projected system axis.
A simple way to explain the elongation and shift is to assume
this is due to superposition of two equally bright and unresolved features separated by 0$\farcs$14, or 20 AU.
\footnote{Strictly speaking, the larger separation in mm emission compared to
optical positions could also be attributed to a shift of the A position, as the
absolute astrometry is not sufficient to decide how to register the two images.
However, shifting the mm emission from A is more difficult because it contributes
a larger fraction of the total mm flux.}
This suggests that B could itself be a close binary, each with a dusty disk emitting
at mm wavelengths.

The long period of decreased optical magnitude for UY Aur B, of duration $D_d \approx 40$
years between approximately 1960 and 2000, would be due to extinction of UY Aur Ba by
the disk of the less massive companion UY Aur Bb. The orbiting period
$\rm {P_{Bb}}$ of the Ba/Bb binary can be estimated from: $\frac{2\pi}{\rm P_{\rm Bb}}=\frac{2\theta}{D_d}$,
where $\theta=\mathrm{asin}(R_b/a)$ and $R_b/a$ is the ratio of Bb disk radius to orbit
semi-major axis $a$, $R_b/a \leq 1/3$ because of tidal truncation. Using $D_d=40$ yrs yields a
period P$_{\rm Bb}$ of 360 yrs. Assuming a total mass B(a+b) of 0.4 M$_{\sun}$ \citep{Hartigan2003},
the separation between Ba and Bb will be 37 AU if the orbit is circular. 
This orbital separation is larger than the deprojected separation of 20 AU from the assumption of equal masses earlier in the section.
Such a discrepancy can be explained if the inclination angle is larger than 42$\degr$. 
If the inclination is up to $\sim$66$\degr$, these two estimated values will be consistent. 
More reasons in favor of larger inclination of the UY Aur system are discussed in Sect. \ref{sec:a_more_inclined_system}.

The flux from each component, around 3.4 mJy each at 1.4\,mm, could
be explained by very compact, $\approx 5$ AU, essentially optically thick disks.  
A spectral index of 1.6-1.8 requires T$<$20 K, which however looks very small for a radius of 5 AU, even though the two stars would be rather low mass objects. 
An alternative is optically thin disks of radii about 20 AU, so that their mean temperature does not exceed 20 K, and with grains large enough to have an emissivity exponent $\beta$ = 0.
This interpretation is thus marginally plausible in terms of sizes and timescales. 
Moreover, in this hypothesis, UY Aur A, Ba and Bb are now nearly aligned with the
projection of the system axis, so an obscuration is only possible if the
orbit of Ba/Bb is perpendicular to the rest of the system.
Yet, the optical jet emanating from B recently detected by
\citet{Pyo+etal_2014} is aligned on the overall system axis which suggests
at least one of the Ba/Bb disk is coplanar with the overall UY Aur system.

\subsubsection{Jets}
\label{sec:discussion:jets}
We already argued in Sect.\ref{sub:discussion:uyaura} that free-free emission
from an optical jet provides an appropriate explanation
to the low spectral index of the mm continuum emission from UY Aur A.
UY Aur B displays a similar spectral index, although the error bar is larger.
It is thus also plausible that the emission from UY Aur B is also the
superposition of dust emission from a CS disk with free-free
emission from an ionized jet. This jet is aligned along
the system axis \citep{Pyo+etal_2014}. 
This can offer a simple
interpretation for both the source elongation and the position offset,
provided only the SE jet emits in free-free. Indeed, \citet{Pyo+etal_2014}
have already shown a strong asymmetry in the [FeII] emission, which
is only detected towards the NW of UY Aur B, while it is detected
on both sides of UY Aur A. Also, an elongated jet offers
more flexibility to explain the relatively large size
of $\sim 0\farcs45$, compared to the simple superposition
of two point sources for which it is difficult to obtain
a size above $\sim 0\farcs3$ for a centroid shift of $0\farcs07$.

\subsection{A more inclined system ?}
\label{sec:a_more_inclined_system}
While the ``disks in a binary'' scenario (Sect. \ref{sec:discussion:uyaurb}) has difficulties explaining
the jet orientation, the ``jet only'' scenario (Sect. \ref{sec:discussion:jets}) unfortunately offers
no explanation for the long duration fadening of UY Aur B. We explore
here a possibility to combine both aspects and  provide a consistent
view of all phenomena affecting UY Aur.

The key difficulty of the binary scenario for UY Aur B is the
near alignment of A, Ba and Bb in combination with the rather low
inclination of the system. On one hand, however,
given the positional uncertainty,
the alignment constraints between the A-B and Ba-Bb lines can be
relaxed to about $45\degr$.
On the other hand, the inclination angle of $42\degr$, which
was first given by \citet{Duvert+etal_1998} is rather uncertain.
\citet{Hioki+etal_2007} quote $42 \pm 3\degr$ from a fit of
an ellipse to the brightest spots in the NIR scattered light image.
However, the foci of the fitted ellipse does not go through the
center of mass of the system, which must be in between A and B,
about twice close from A than from B. \citet{Hioki+etal_2007}
also pointed out that the phase function of scattering cannot
be anisotropic enough to explain the NW/SE contrast with this
inclination and some asymmetric extinction is needed to explain
northern-southern difference.

 The dynamical mass of $1.2~\Msun$ estimated by
\citet{Duvert+etal_1998} for $i=42^\circ$ becomes
$0.7~\Msun$ for $i= 60\degr$. This is small compared
to the latest estimate of the stellar masses from the spectral
type by \citet{Hartigan2003}, being 0.60 and $0.34~\Msun$ for A
and B respectively. We note that the identification of a Keplerian
pattern in the emission from the CO isotopologues is difficult,
so the dynamical mass is rather uncertain. However, \citet{Hioki+etal_2007}
derived a total binary mass to be $1.73\pm 0.29~\Msun$ using the $42^\circ$
inclination: this is already above the $0.94~\Msun$ value,
and should scale as $1/\cos^3(i)$ since the A-B separation is
nearly aligned with the projected minor axis. A higher inclination
can thus be accommodated only if the orbit is elliptical and the stars
near apastron.

Hence, a higher inclination of $\sim 50\degr$ to 60$\degr$ for the overall system,
combined with inclination differences up to $10\degr$ between
the various disks and orbital planes, may be sufficient to allow
obscuration of Ba by Bb. The stability of such a system has also
to be investigated. Furthermore, while Ba is exiting the occultation by Bb, its southern-eastern jet may still be behind the Bb disk,
offering a direct explanation for the lack of detection of
[FeII] in this direction by \citet{Pyo+etal_2014}.

We note that these differences in inclination are reminiscent
of the situation in AB Aur, where the presence of surrounding
material leads to an apparent inclination  quite different
from that of the inner disk \citep{Tang+etal_2012,Pietu+etal_2005}.
The surroundings of UY Aur are complex and may
still influence the evolution of the system.

\begin{figure*}[ht!]
\includegraphics[scale=0.5]{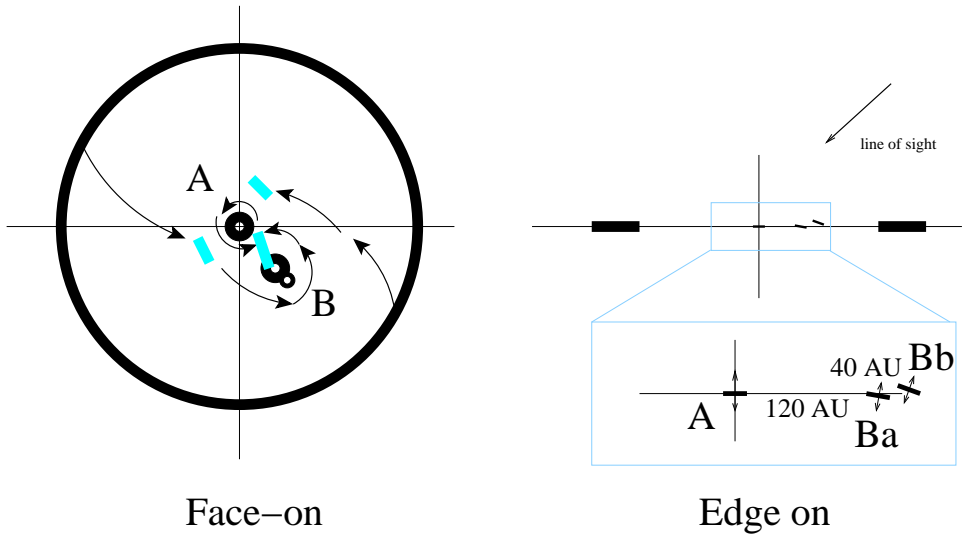}
\caption{Schematics of the UY Aur system in the face-on view and in the edge-on view. 
In the face-on view, the cyan segments denote the shock region traced with the SO line.
The arrows mark the flows between the CB disk (shown in large-black circle) and the CS disks (shown in small-solid-black circles).
In the edge-on view, the segments mark the CS disks, and the arrows mark the outflows/jets.
}\label{fig:schematics}
\end{figure*}

Figure \ref{fig:schematics} is a scheme which summarizes most of the observed
properties of the UY Aur system.
The accreting material between the CB and CS disks is better demonstrated in the position-velocity diagrams (Figure \ref{fig:pv}).
The proposed scenario (binary) can be tested by searching for a binary
companion (Ba) at 40 AU from B (Ba) using NIR interferometric techniques.
Similar result has recently shown by \citet{DiFolco+2014} who found that GG
Tau A is in fact a triple system, the second binary located around
GG Tau Ab having a separation of 5 AU.
If the discussed scenario is valid, the UY Aur should be a triple system.

\section{Conclusion and Summary}
Using the IRAM array and the SMA, we have studied the properties of the molecular gas and
dust orbiting around the binary system UY Aur. We observed the circimbinary (CB) ring, the streamers and
the circumstellar (CS) disks with the molecular lines $^{13}$CO and C$^{18}$O 2-1, SO 5(6)-4(5) and $^{12}$CO 3-2.
The thermal dust emission at 3.4 mm, 1.4 mm and 0.85 mm is also studied. The results are summarized below:

\begin{enumerate}
\item The gaseous component of the CB ring is easily detected in $^{13}$CO and C$^{18}$O but
its dust counterpart, even at 1mm continuum, is very marginally seen. We derive a total mass of
 $M_{\rm gas} \sim 2 \times 10^{-3}$ M$_{\sun}$.
A gaseous component associated with NIR 1 in the south-east of the CB ring is detected, which is probably in-falling toward the CB ring.

\item The inflowing cold gas emanating from the CB ring toward the CS disks is clearly detected in $^{13}$CO and C$^{18}$O
following spiral patterns in the NE and in the SW of the CB disk, revealing that accretion proceeds through streamers.
Combinations of a Keplerian CB ring, inflowing gas and also probably outflows make a complex velocity pattern and precludes
any accurate determination of the mass of the central binary.

\item The detections of $^{12}$CO 3-2 and SO in the vicinity of the CS disks suggest gradients of physical conditions
in the streamers and likely accretion shocks, at least near the CS disk of UY Aur A. This scenario is also
supported by the detection of bright $H_{2}$ patches near UY Aur A \citep{Beck+inprep}.

\item At least, two CS disks are well detected in the mm continuum emission.
The spectral indices $\alpha$ are 1.62$\pm$0.1 and 1.62$\pm$0.3 for UY Aur A and B, respectively.
Such low spectral indices are more likely due to the combination of dust emission and
free-free continuum emission.

\item To reconcile the optical variability of UY Aur A with the existing bright mm dust disk, we
propose the object orbiting UY Aur A in close orbit ($6$\,AU) is a planet which
mass is not enough to open a large gap in the disk around UY Aur A.

\item To reconcile the optical and Infrared variability of UY Aur B, we propose this
object is a binary of separation of the order of 40 AU (to take into account the long duration
of the obscuration $\sim$ 40 yrs). Such a binary scenario is compatible with the presence of a jet
but requires a higher inclination for the overall system (55-60$^o$ instead of $42^o$),
combined with inclination differences as high as 10$^o$ between the various disks and orbital planes.

\end{enumerate}

In conclusion, these observations reveal that, if accretion proceeds through streamers in binary systems,
the streamer configurations, their strengths and their dynamics are difficult to constrain.
They have to be investigated in detail to confront models to the observations, requiring
the sensitivity and resolving power of ALMA complemented by optical (interferometric) observations.\\

{\it Facilities:} \facility{PdBI}, \facility{SMA}.

\acknowledgments{We acknowledge the IRAM staff at Plateau de Bure and Grenoble and the SMA staff for carrying out the observations. 
This research was partially supported by NSC grants NSC99-2119-M-001-002-MY4.
This research was partially supported by PCMI, the French national program for
the Physics and Chemistry of the Interstellar Medium. 
This research has made use of 
the SIMBAD database, operated at CDS, Strasbourg, France, and of the NASA ADS Abstract Services.}
%
%
\bibliographystyle{apj}                       
\bibliography{uyaur2}

\end{document}